\documentclass[aps,pre,floatfix,reprint,groupedaddress,showkeys,superscriptaddress]{revtex4-1}
\usepackage[utf8x]{inputenc}
\usepackage{graphicx}
\usepackage{dcolumn}
\usepackage{bm}
\usepackage{epstopdf}
\usepackage{amsmath}
\usepackage{float}
   \usepackage[all]{xy}
\usepackage{subfigure}
\usepackage{xcolor}
\usepackage{soul} 

\begin{document}
	
\title{Statistical physics of interacting proteins:\\ impact of dataset size and quality assessed in synthetic sequences}

 \author{Carlos A. Gandarilla-Pérez}
\affiliation{Sorbonne Universit{\'e}, CNRS, Institut de Biologie Paris-Seine, Laboratoire de Biologie Computationnelle et Quantitative (LCQB, UMR 7238), F-75005 Paris, France}
 \affiliation{Facultad de Física, Universidad de la Habana, San Lázaro y L, Vedado, Habana 4, CP–10400, Cuba}
 
 \author{Pierre Mergny}
 \affiliation{Sorbonne Universit{\'e}, CNRS, Institut de Biologie Paris-Seine, Laboratoire de Biologie Computationnelle et Quantitative (LCQB, UMR 7238), F-75005 Paris, France}  
 \affiliation{Sorbonne Universit{\'e}, CNRS, Institut de Biologie Paris-Seine, Laboratoire Jean Perrin (LJP, UMR 8237), F-75005 Paris, France}  
 
 \author{Martin Weigt}
 \email[Corresponding author: ]{martin.weigt@sorbonne-universite.fr}
 \affiliation{Sorbonne Universit{\'e}, CNRS, Institut de Biologie Paris-Seine, Laboratoire de Biologie Computationnelle et Quantitative (LCQB, UMR 7238), F-75005 Paris, France}
 
  \author{Anne-Florence Bitbol}
  \email[Corresponding author: ]{anne-florence.bitbol@epfl.ch}
  \affiliation{Sorbonne Universit{\'e}, CNRS, Institut de Biologie Paris-Seine, Laboratoire Jean Perrin (LJP, UMR 8237), F-75005 Paris, France}
  \affiliation{Institute of Bioengineering, School of Life Sciences, École Polytechnique Fédérale de Lausanne (EPFL), CH-1015 Lausanne, Switzerland}

\begin{abstract}
Identifying protein-protein interactions is crucial for a systems-level understanding of the cell. Recently, algorithms based on inverse statistical physics, e.g. Direct Coupling Analysis (DCA), have allowed to use evolutionarily related sequences to address two conceptually related inference tasks: finding pairs of interacting proteins, and identifying pairs of residues which form contacts between interacting proteins. Here we address two underlying questions: How are the performances of both inference tasks related? How does performance depend on dataset size and the quality? To this end, we formalize both tasks using Ising models defined over stochastic block models, with individual blocks representing single proteins, and inter-block couplings protein-protein interactions; controlled synthetic sequence data are generated by Monte-Carlo simulations. We show that DCA is able to address both inference tasks accurately when sufficiently large training sets of known interaction partners are available, and that an iterative pairing algorithm (IPA) allows to make predictions even without a training set. Noise in the training data deteriorates performance. In both tasks we find a quadratic scaling relating dataset quality and size that is consistent with noise adding in square-root fashion and signal adding linearly when increasing the dataset. This implies that it is generally good to incorporate more data even if its quality is imperfect, thereby shedding light on the empirically observed performance of DCA applied to natural protein sequences.
\end{abstract}

\maketitle

\section{Introduction}

Most cellular processes are carried out by interacting proteins, and mapping functional protein-protein interactions is a crucial question in biology. Since genome-wide experiments remain challenging~\cite{Rajagopala14}, an attractive possibility is to directly exploit rapidly expanding sequence databases in order to identify protein-protein interaction partners.

Methods inspired by inverse statistical physics~\cite{nguyen2017inverse} have recently received increasing interest in computational protein biology, as reviewed in~\cite{Cocco18}. The basic idea is simple: in the course of evolution, proteins diversify considerably their amino-acid sequences, while keeping their three-dimensional structure, their biological function and, most importantly in our context, their inter-protein interactions remarkably well conserved. Families of homologous proteins, i.e.~proteins of common evolutionary ancestry, therefore offer samples of highly variable sequences of common structure and function. In many cases, these families contain $10^3-10^6$ distinct sequences, and statistical approaches are well adapted to analyze sequence variability, and to unveil hidden information about the proteins' behavior and the selective forces acting on sequence evolution.

In the case of individual protein families, global statistical models~\cite{Lapedes99} based on the maximum-entropy principle~\cite{Jaynes57} and assuming pairwise interactions, known as Direct Coupling Analysis (DCA)~\cite{Weigt09,Morcos11}, PsiCOV~\cite{jones2011psicov} or GREMLIN~\cite{balakrishnan2011learning}, have been used with success to predict three-dimensional protein structures from sequences~\cite{Marks11,ovchinnikov2017protein}, to analyze mutational effects~\cite{Lui13,Dwyer13,Cheng16,Figliuzzi16,hopf2017mutation,levy2017potts} and conformational changes~\cite{Morcos13,Malinverni15}.

In the more complex case of interacting proteins, two protein families are investigated jointly, cf.~\cite{szurmant2018inter} for a recent review. Inference is based on the idea that the amino-acid sequences of interacting proteins are correlated, in particular because contacting amino acids need to maintain physico-chemical complementarity through evolution. At least three questions can be asked:
\begin{itemize}
\item Do the two families interact, i.e.~is there a substantial number of interacting protein pairs with the two interaction partners belonging to the two protein families? DCA-based approaches have been proposed, detecting interaction via the strength of statistical couplings between protein families~\cite{Feinauer16,malinverni2017modeling,croce2019multi,cong2019protein}.
\item Which specific proteins from the two families interact? While this problem can be partially solved by pairing proteins that exist in the same species (i.e. the same genome), a given genome often contains several homologous members (called paralogs) of each of the two protein families, thus making interaction-partner matching a non-trivial problem. Global statistical models~\cite{Burger08}, and in particular DCA, can address this question both in the cases when a substantial~\cite{Procaccini11} or small~\cite{Gueudre16} training set of known interaction partners is available, and even in the absence of such a training set~\cite{Bitbol16}. Iterative algorithms have been developed in the last two cases~\cite{Bitbol16,Gueudre16}. In practice, training sets may correspond to pairs of interacting partners known from experiments or from genomic colocalization~\cite{Procaccini11}, but in cases where it is not known whether the two protein families considered interact or not, there is no training set.
\item How do these proteins interact, i.e. which residues form the interaction interface? DCA-type approaches have helped in finding residue contacts between known interaction partners and in protein-complex assembly~\cite{Weigt09,Ovchinnikov14,Hopf14,Tamir14,dosSantos15,malinverni2017modeling}.
\end{itemize}
While promising applications have been presented for all three questions, theoretical problems related to the underlying inference procedures remain open. Here we address the last two questions, which involve two coupled inference tasks: inferring interaction partners among paralogs from sequence data, and inferring contacts between amino acids. How do these two tasks couple together? What is the impact of dataset size and quality on their performance? 

\begin{figure}[h!]
	\begin{center}
		\includegraphics[keepaspectratio,width=\columnwidth]{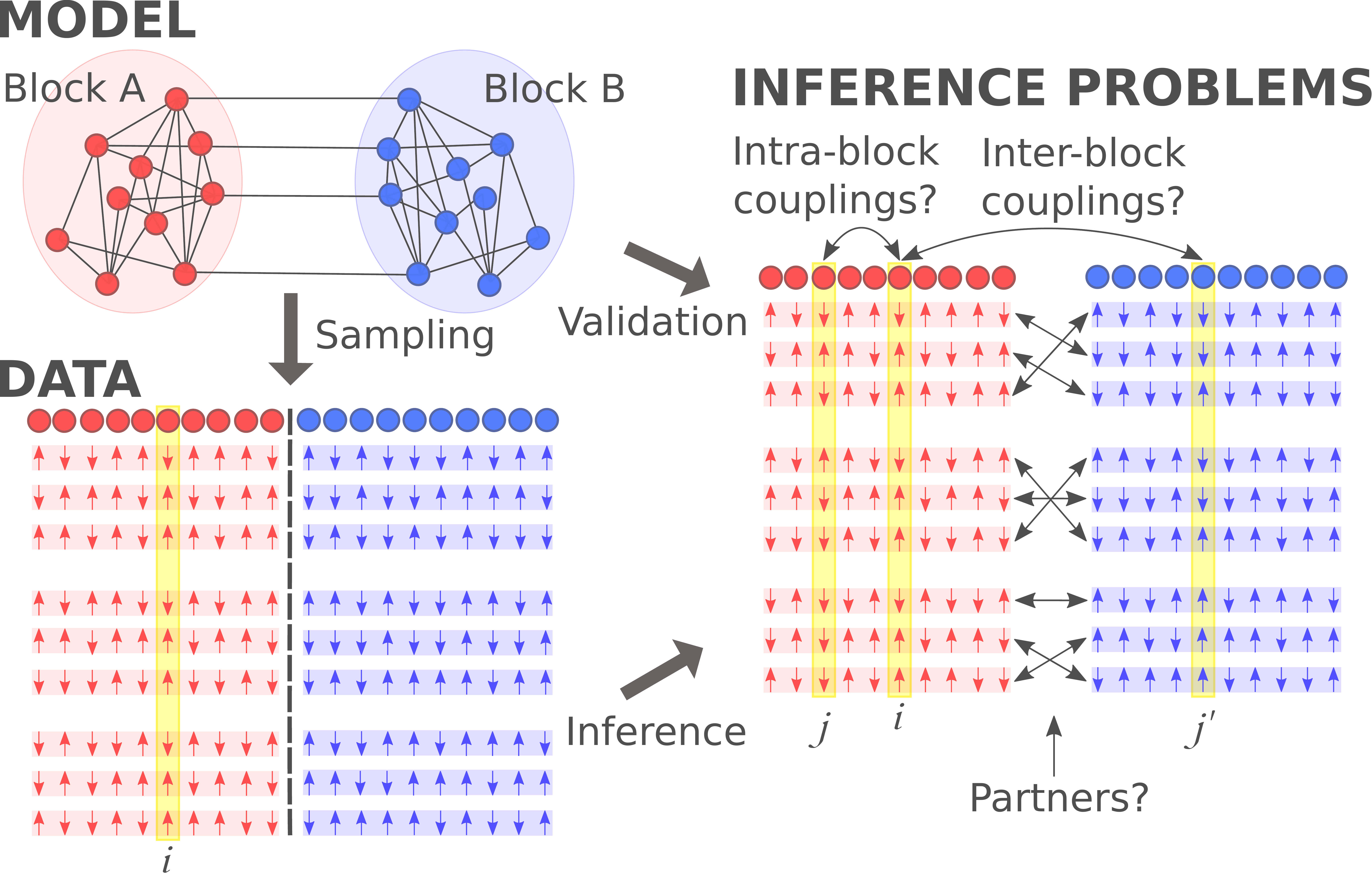}
	\end{center}
	\vspace{-5mm}
	\caption{Schematic illustration of our model, the data generation procedure and the inference tasks: Interacting proteins are represented as blocks in a stochastic block model. Data are generated by equilibrium Monte-Carlo simulations of an Ising model defined on the SBM graph; they are subdivided into groups mimicking species, and split into half chains A and B corresponding to individual SBM blocks and representing single proteins. The original pairing between chains A and B is blinded. The resulting inference tasks are the reconstructing of the inter-block couplings (predicting residue contacts between proteins) and the pairing of chains A and B (predicting interacting protein pairs).} 
        \label{graphical_abstract}
\end{figure}

To address these questions, we propose the well controlled setting for synthetic data generation schematically represented in Fig.~\ref{graphical_abstract}. The interacting proteins are mimicked by a stochastic block model (SBM)~\cite{dyer1989solution}, a well-known random graph model used to represent modular networks by incorporating blocks with different connectivities inside the blocks and between them. Concretely, there is a certain number of internal couplings in each block, and there are a certain number of inter-block couplings. Here, we consider two blocks, representing the two interacting proteins: vertices represent amino acids, while edges represent either intra-protein or inter-protein couplings, depending on their location inside or between proteins. Data are generated by equilibrium Monte Carlo simulations according to an Ising Hamiltonian with identical couplings on all edges of the SBM graph, cf.~details in App.~\ref{App_data_gen}. 

To mimic the situation found in real proteins, spin chains are randomly divided into groups (representing species), and split into two halves (chains A and B, representing proteins) according to the two blocks, cf.~Fig.~\ref{graphical_abstract}. We next blind the pairings between halves, i.e.~a spin chain A could in principle be paired with any of the spin chain B inside the same group. This represents the fact that interaction partners have to belong to the same species, and that within a species, pairings among paralogs are not a priori known. The sets of all sampled chains A and of all sampled chains B each represent a multiple-sequence alignment (MSA) of a protein family. Based on this construction, our two inference tasks can be formalized:
\begin{itemize}
\item Can we infer the couplings, or edges, between the two blocks of the original SBM graph from data?
\item Can we determine the correct pairing between the spin chains A and B in the blinded data?
\end{itemize}
We start by addressing these problems in the presence of a training set, i.e.~an ensemble of paired chains A and B generated together. With natural data, ideally, the training set contains only correct pairs of interacting partners, but in reality there may be incorrect pairs in the training set, e.g. due to experimental challenges~\cite{Rajagopala14}. Therefore, here, we explicitly address the realistic case where some pairings in the training set are incorrect, i.e. associate two independently generated chains A and B, which introduces noise into the inference problem. Next, we consider the case without a training set.

The paper is organized as follows. Sec.~\ref{sec_model} briefly defines the model and the data generating process, with details in the Appendices. In Sec.~\ref{sec_couplings}, we address the problem of inferring inter-block couplings, and consider the impact of training set size and quality. A similar analysis for the partner prediction problem follows in Sec.~\ref{sec_partners}. To quantify the impact of training set size and quality, we present a scaling analysis of the influence of noise in the training set in Sec.~\ref{sec_scaling}. The case without a training set is presented in Sec.~\ref{sec_noTS}. Finally, conclusions are presented in Sec.~\ref{Sec_ccl}.

\section{Model and data generation}
\label{sec_model}

The model illustrated in Fig.~\ref{graphical_abstract} is defined over an SBM graph having two blocks A and B, each one with $L=100$ vertices, yielding a total of 200 vertices. Any two vertices inside a block are connected with probability $p_{intra}=0.025$, while any two vertices belonging to different blocks are connected with $p_{inter}=0.02$. We have checked that the qualitative results do not depend on these specific values, cf.~App.~\ref{app_length} for the case of smaller blocks of size $L=50$.

In addition to this graph, we define a ferromagnetic model having an Ising spin on each vertex, and equal couplings on all edges of the graph. It is defined by its Hamiltonian
\begin{equation}
  H(\vec \sigma) = - \sum_{(ij)\in {\cal E}} \sigma_i \sigma_j\; ,
\end{equation}
where $\vec \sigma = (\sigma_1,...,\sigma_{2L})\in \{\pm 1\}^{2L}$ is a vector of $2L$ Ising spins, and ${\cal E}$ the set of edges of the SBM graph. We will denote $(\sigma_1,...,\sigma_{2L})$ as chain AB, and the halves $(\sigma_1,...,\sigma_{L})$ (resp. $(\sigma_{L+1},...,\sigma_{2L})$) corresponding to block A (resp. B) as chain A (resp. B). 

Data are generated from this model using Monte Carlo simulations, see App.~\ref{App_data_gen} for details. The sampling scheme was designed to obtain spin chains corresponding to identically and independently distributed equilibrium spin chains at a formal temperature slightly above the ferromagnetic phase transition temperature of our graph (see App.~\ref{app_length}). We have also studied the impact of varying sampling temperature (see Fig.~\ref{PPVinter_vs_T_N100-200}), showing that good performance is also obtained in the ferromagnetic phase slightly below the phase transition. For sample averaging, we generated a sample of 150,000 equilibrium spin chains for one specific SBM graph, and randomly extracted disjoint training and testing sets from this large sample. We did not average over different realizations of the graph, since in the protein case we are interested in the influence of dataset size and quality for a given pair of interacting protein families, i.e.~the model does not change, but the data do. Because inter-protein contacts are sparse, and often sparser than intra-protein ones, we have further compared the results obtained with our baseline graph to those obtained with variants of this graph, namely one with lower inter-block connectivity and one with inter-block couplings restricted to a subset of sites of each block (see App.~\ref{connectivities_app}, Fig.~\ref{Connectivities}).

Compared to natural protein sequences, our synthetic data is idealized, because it is fully at equilibrium, and all correlations are from actual couplings, excluding any impact of phylogeny. This is because our aim here is to quantitatively understand the bases of prediction of interaction partners thanks to contacts. Correlations arising in protein sequences due to their common evolutionary history~\cite{Casari95,Halabi09,Qin18,Fryxell96,Goh00,Pazos01,Ochoa10} can further contribute to the success of DCA-based approaches at predicting protein-protein interactions from natural protein sequences~\cite{BitbolPMI,Marmier19}, while they obscure the identification of contacts~\cite{Weigt09,Marks11,Qin18}. Besides, generalizing the present model to Potts spins and including local fields to tune residue conservation would make it more similar to the models inferred by DCA from natural protein sequences. One could also go beyond our simple ferromagnetic model and test the impact of the heterogeneity of interactions, since matching residue conservation in synthetic data with that of natural protein sequences has shown that protein alignments may lie in the frozen phase of a frustrated system~\cite{Franco19}. More generally, the question of which phase is relevant in natural data is a very interesting one, with several studies suggesting that natural data may be close to criticality~\cite{Mora11}. In this paper, we have chosen to use the simplest non-trivial model to address the coupled inference problems of finding interacting spins and interacting chains, aiming to extract quantitative laws in a well-controlled case.

\section{Inference of couplings}
\label{sec_couplings}

\begin{figure}[h!]
	\begin{center}
		\includegraphics[keepaspectratio,width=\columnwidth]{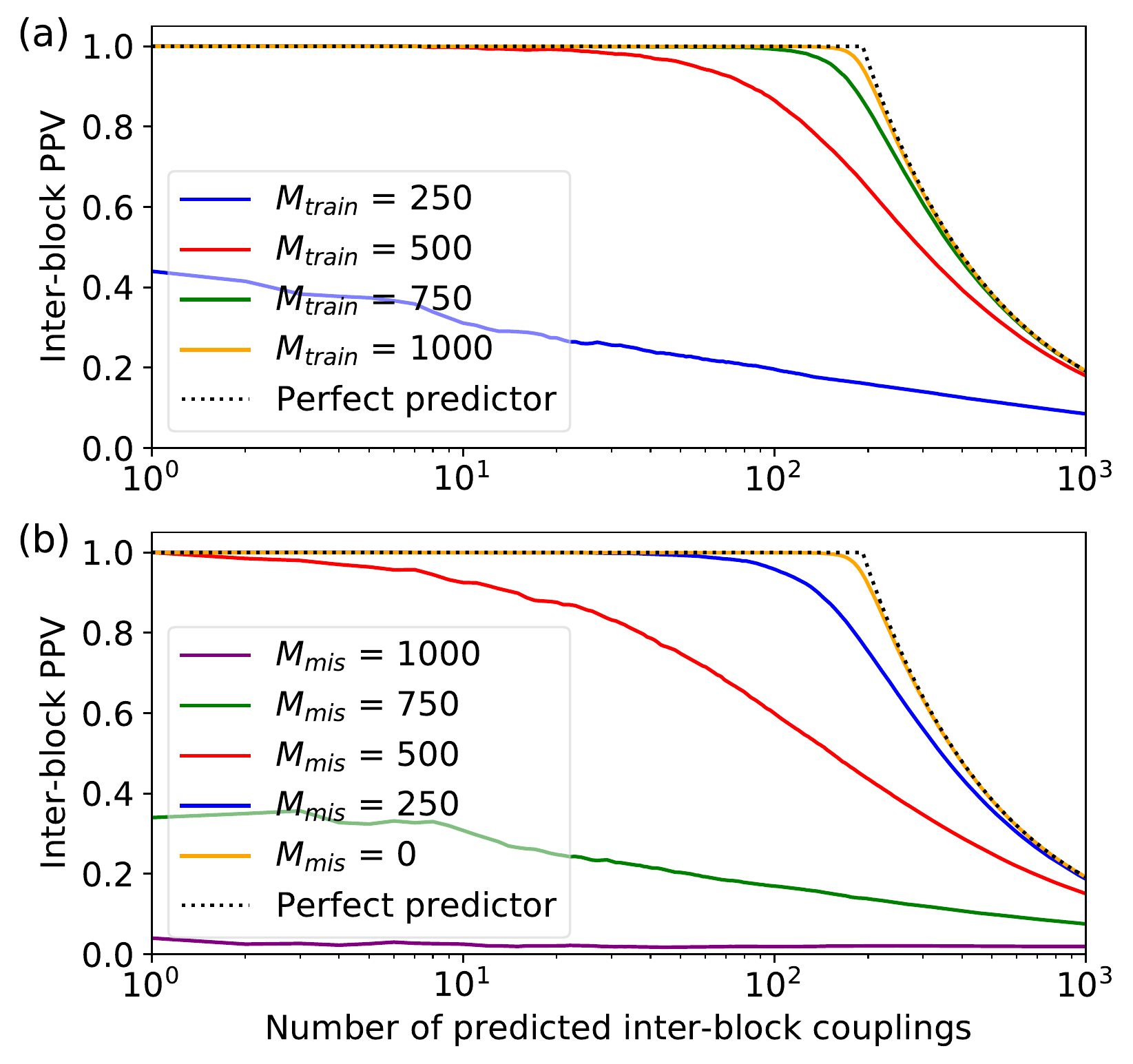}
	\end{center}
	\vspace{-5mm}
	\caption{Positive predictive value (PPV) of the inferred inter-block couplings, i.e. fraction of the top inferred inter-block couplings that are actual inter-block couplings in the stochastic block model (SBM) random graph used to generate the data, plotted versus the number of top inter-block couplings considered. Data is generated using a SBM(0.025,0.02) graph with $L = 100$ spins per single chain A or B at a sampling temperature $T = 5.0$. We then randomly extract (without replacement) a training set from the total generated dataset, and DCA inference is performed. Averages over 100 realizations corresponding to different training sets are shown. (a): Training sets with no mismatches comprising different numbers $M_{train}$ of chains AB. (b): Training sets with total number $M_{train}=1000$ of chains AB, comprising different numbers $M_{mis}$ of mismatched pairs AB. }\label{ppv_inter}
\end{figure}

Consider the first inference problem, namely that of couplings between spin sites. To address it, we first apply DCA~\cite{Weigt09,Cocco18} within the mean-field approximation~\cite{Plefka82,Morcos11} to infer a Hamiltonian
\begin{equation}
  H_{DCA}(\vec \sigma) = - \sum_{i<j} J_{ij} \sigma_i \sigma_j\; ,
\end{equation}
which a priori contains couplings $J_{ij}$ for each pair $(i,j)$ of sites. This simple mean-field approximation has been used with success for protein structure prediction~\cite{Morcos11,Marks11} and is computationally fast, which is important in our framework since the determination of interaction partners in natural sequence data involves iterative procedures where the inference task is performed multiple times~\cite{Bitbol16,Gueudre16}.  Next, we ask whether the top inferred couplings $J_{ij}$ correspond to actual couplings within the model, i.e.~to edges of the underlying SBM graph (see App.~\ref{App_infer} for details on this inference method). We distinguish the case of intra-block couplings and of inter-block ones, with the intuition that the blinding of pairings between chains A and chains B within each species may obscure the inference of inter-block couplings more than that of intra-block ones. Fig.~\ref{ppv_inter} focuses on inter-block couplings, and shows the positive predictive value (PPV) of the inferred couplings, and Fig.~\ref{ppv_intra} shows a similar analysis in the case of intra-block couplings.  Note that because we are working with Ising spins, PPV is defined directly with the inferred couplings $J_{ij}$, without the need to compress the data obtained with each amino-acid types for a given pair of sites $(i,j)$. For natural sequences, this compression step is usually done using Direct Information~\cite{Weigt09} or Frobenius norm~\cite{Ekeberg13}. 

We first consider the case where chains AB are correctly paired in the training set used to construct the DCA model (see App.~\ref{App_infer}). Fig.~\ref{ppv_inter}(a) shows that it is necessary to have a large enough training set ($M_{train}\sim500$ chains AB) in order to obtain good performance on inter-block couplings, and that performance is almost perfect for a training set of $M_{train}\sim1000$ chains AB. Although the exact training set size required for good performance will depend on the number of states per site and on the number of sites, the order of magnitude is consistent with the (effective) number of sufficiently different protein sequences needed for DCA to infer contacts between amino acids~\cite{Weigt09,Morcos11,Marks11}. Fig.~\ref{ppv_intra}(a) shows very similar results in the case of intra-block couplings. Indeed, with perfectly paired chains AB in the training set, we do not expect inter- and intra-block couplings to behave differently.

Next, we investigate the impact of having mismatches between chains A and B in the training set, which would occur in natural data if the interaction partners were not or imperfectly known to begin with. Fig.~\ref{ppv_inter}(b), which employs a training set of $M_{train}=1000$ chains AB including various numbers $M_{mis}$ of mismatched pairs, shows that mismatches deteriorate the inference of inter-block couplings. Conversely, we observe no deterioration of the inference of intra-block couplings due to the mismatches in Fig.~\ref{ppv_intra}(b).

In section~\ref{sec_scaling}, we will investigate in more detail the impact of both (training) dataset size and quality, and demonstrate how performance scales with them, both for the inference of couplings (present problem) and for the inference of partners (next problem).

\section{Inference of partner pairings}
\label{sec_partners}

Let us now address the second inference problem, namely that of inferring partner chains AB in a testing set where partnerships are blinded, from a training set where partnerships are known. By ``partner chains'' we mean pairs that come from the same complete chain AB generated from the the stochastic block model. To mimic the realistic case where possible interacting partners among proteins would consist of proteins within the same species, we use the before-mentioned random split of our testing set into groups, here with 10 pairs AB each, and then we blind the pairings within each of these groups, cf.~Fig.~\ref{graphical_abstract}.

\begin{figure}[htb]
	\begin{center}
		\includegraphics[keepaspectratio,width=\columnwidth]{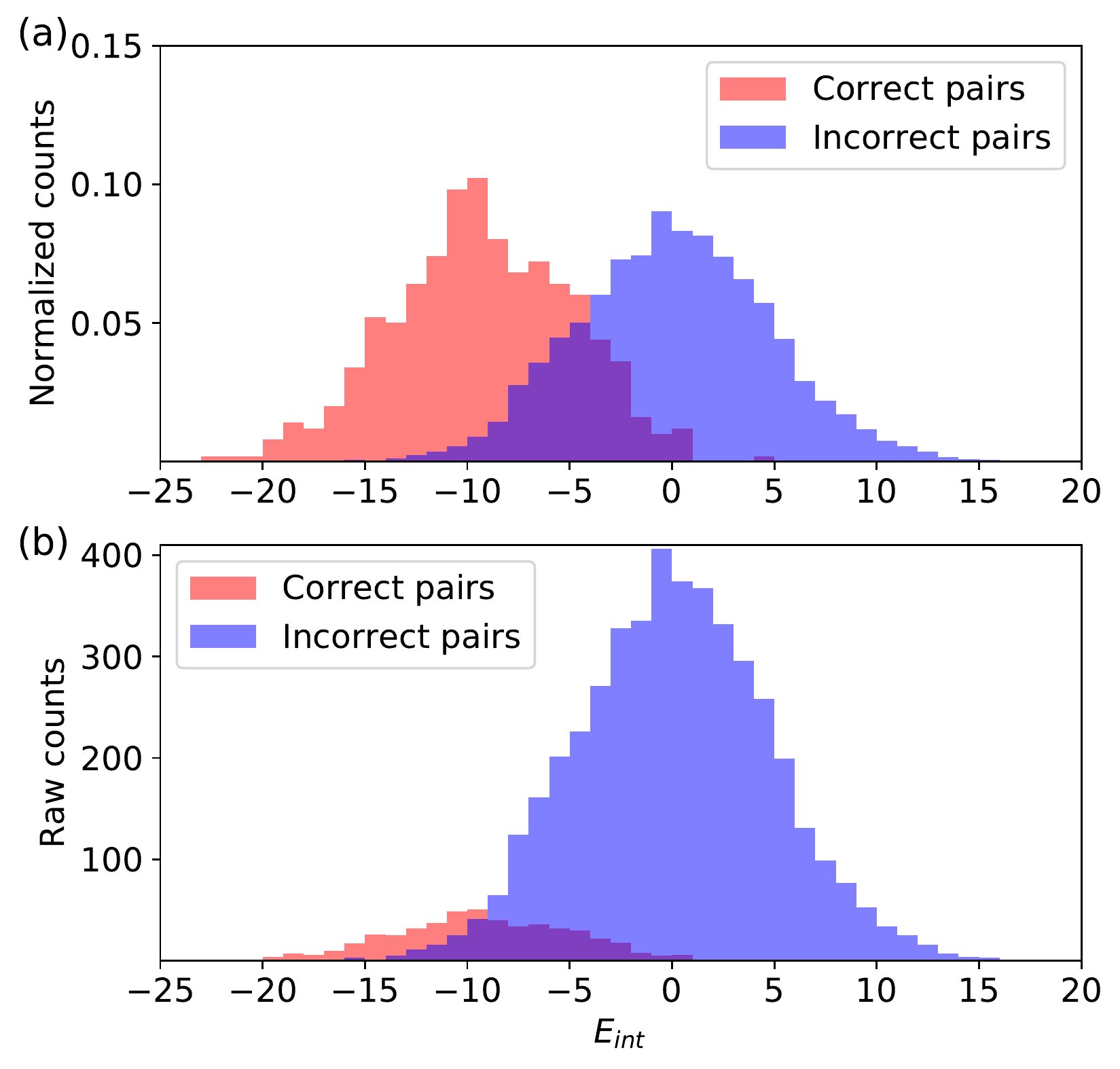}
	\end{center}
	\vspace{-5mm}
	\caption{Histograms of the interaction energies $E_{int}$ (see Eq.~\ref{Eint_main}) for correct and incorrect within-species pairs AB in the testing set. The couplings in Eq.~\ref{Eint_main} are computed employing a training set where all pairs AB are correct (no mismatches). (a): Normalized histograms. (b): Raw count histograms (same data). The testing and training sets respectively comprise $M_{test} = 500$ and $M_{train} = 2000$ spin chains AB. In the testing set, species comprising 10 pairs AB each are randomly constructed, and matchings are then blinded within each species. Data is generated using the same SBM(0.025,0.02) graph with $L = 100$ spins per single chain A or B as in Fig.~\ref{ppv_inter}, also at a sampling temperature $T = 5.0$. }\label{Eint_histogram}
\end{figure}

Given that the correct pairs were generated by equilibrium sampling from the Hamiltonian associated to an SBM graph, we expect that they will feature smaller interaction energies 
\begin{equation}
E_{int}=-\sum_{i=1}^{L}\sum_{j=L+1}^{2L} J_{ij}\sigma_i^A\sigma_j^B\,.
\label{Eint_main}
\end{equation}
than incorrect ones (note that the $J_{ij}$ employed are the inferred couplings). This idea has been used with success to predict interaction partners among paralogous proteins in ubiquitous prokaryotic protein families~\cite{Procaccini11,Bitbol16,Gueudre16}. Fig.~\ref{Eint_histogram}(a) shows that this is indeed the case on average, but there is overlap between the observed distributions of interaction energies of correct and incorrect pairs AB. This is problematic when using $E_{int}$ as a pairing score, in particular since a species with $k$ chains AB allows for $k$ correct pairings, but also for $k(k-1)$ incorrect ones. In the raw histograms of Fig. \ref{Eint_histogram}(b), the overlapping region therefore becomes dominated by the more abundant incorrect pairings. Note that this issue will increase with the number $k$ of chains AB per species (Fig.~\ref{Eint_histogram}(b) uses $k=10$).

\begin{figure}[h!]
	\begin{center}
		\includegraphics[keepaspectratio,width=\columnwidth]{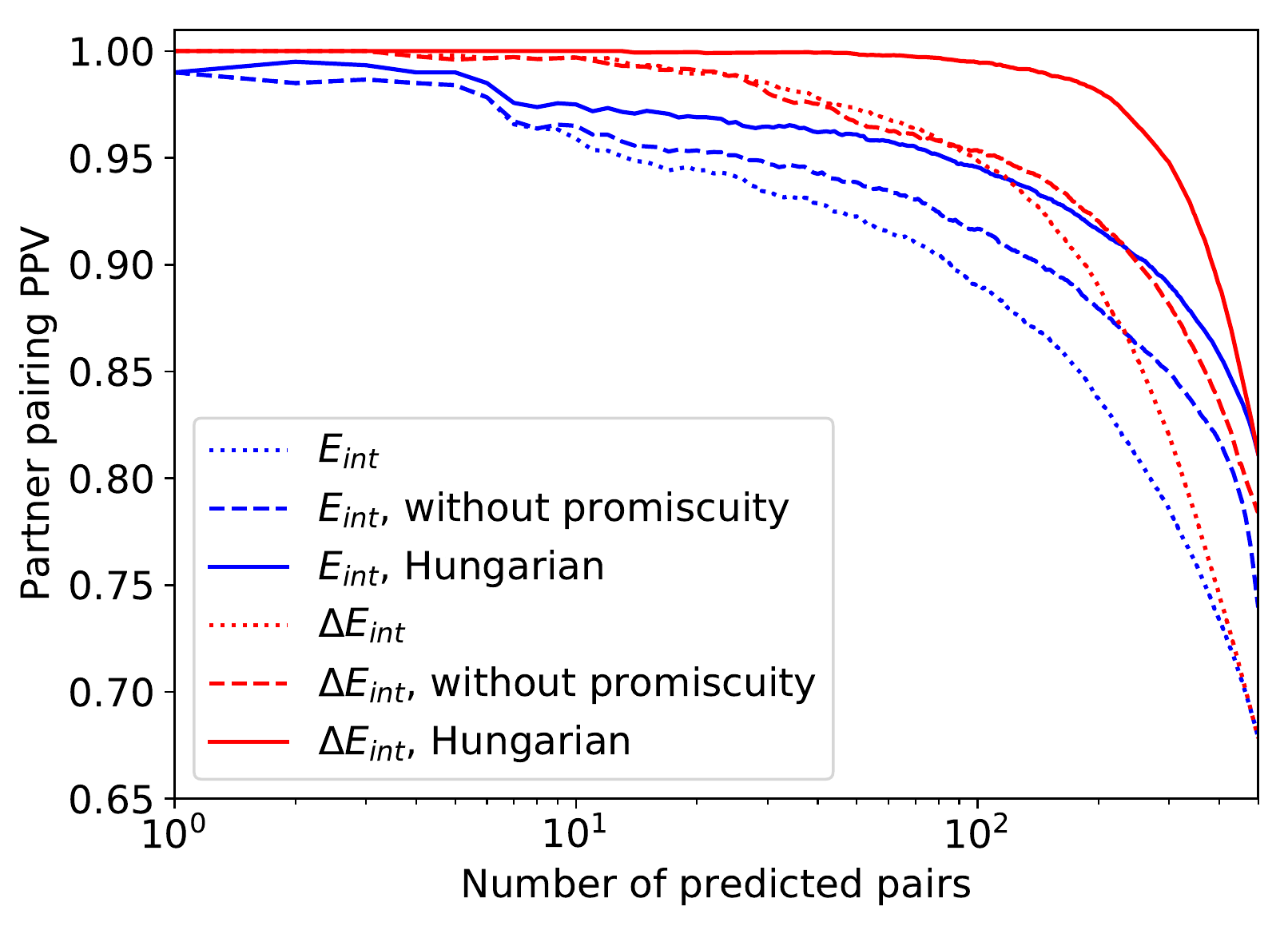}
	\end{center}
	\vspace{-5mm}
	\caption{Positive predictive value (PPV) of pairing prediction, i.e. fraction of the top ranked predicted pairs of chains AB that are correct partners, plotted versus the number of top predicted pairings considered. Data is generated using the same SBM(0.025,0.02) graph with $L = 100$ spins per single chain A or B as in Fig.~\ref{ppv_inter}, also at a sampling temperature $T = 5.0$. We then randomly extract (without replacement) a training set and a testing set from the total generated dataset, comprising $M_{train} = 2000$ and $M_{test} = 500$ complete chains AB respectively. The DCA Hamiltonian is inferred on the spin chains AB of the training set, where no mismatches are introduced. Next, the chains in the testing set are randomly partitioned into species with 10 pairs AB each, and within each species, pairings are blinded. Pairs AB are then predicted within each species using the interaction energy $E_{int}$, either allowing multiple B chains to be matched with the same A (``promiscuity''), or not, or employing the Hungarian algorithm that minimizes the sum of interaction energies.  Finally, predicted pairs are ranked using two different confidence scores, namely the interaction energy $E_{int}$ and the energy gap $\Delta E_{int}$. Averages over 100 realizations corresponding to different training and testing sets are shown.}\label{ppv_partner_H12}
\end{figure}

Starting from the idea that pairs AB with smaller values of $E_{int}$ are more likely to be correct partners, there are several ways to make predictions. The first one is to simply take the chain B that has the smallest value of $E_{int}$ for each chain A. Another one is to include the hypothesis that matchings are one-to-one, and to therefore disallow matching several A chains with the same B chain and vice-versa. This can be done in a greedy way within each species, first taking the pair AB with the lowest $E_{int}$ in the species, then suppressing the corresponding A and B from further consideration, and moving on to the next lowest $E_{int}$ in the species, until all A and B chains are matched~\cite{Bitbol16}. Another possibility is to find the one-to-one assignment that minimizes the sum of all interaction energies $E_{int}$ within the species~\cite{BitbolPMI,Marmier19}, i.e.~to construct the overall optimal matching. This can be done exactly using the Hungarian algorithm (also known as the Munkres algorithm)~\cite{Kuhn55,Munkres57,HungAlg}. Fig.~\ref{ppv_partner_H12} demonstrates that the performance of matching prediction increase when moving on from the first method to the second one, and from the second one to the third one.

Because of the overlap of the distributions of interaction energies $E_{int}$ (see Fig.~\ref{Eint_histogram}), the comparison of $E_{int}$ values does not ensure perfect predictions of correct partners. Therefore, it is also useful to assess confidence in the predictions made employing $E_{int}$ values. The simplest way to do this is to rank them by increasing order of $E_{int}$. In Ref.~\cite{Bitbol16}, a more sophisticated confidence score based on the energy gap $\Delta E_{int}$, namely the difference of interaction energies between the predicted pair and the next best possible pair involving the same chain A, was successfully employed on natural sequence data. Note that in Ref.~\cite{Bitbol16}, the energy gap was corrected to incorporate the fact that matchings are easier in species with fewer pairs of sequences, yielding better performance, but this confidence score reduces to the gap in the simplified setting we consider here, since all our species contain the same number of pairs of sequences. When using the Hungarian algorithm, the associated energy gap score is defined in a global way for each predicted pair, by using the difference of scores between the optimal assignment of pairs in the species and the best alternative assignment that does not involve this predicted pair~\cite{BitbolPMI}. Note that for the natural data considered in Refs.~\cite{Bitbol16,BitbolPMI}, the use of this corrected gap score together with the greedy algorithm slightly outperformed the use of the Hungarian algorithm with the associated gap in the case of DCA inference~\cite{BitbolPMI}. 

Thus motivated, we tested the performance of the energy gap as a confidence score. Fig. \ref{ppv_partner_H12} confirms that the energy gap is a more efficient score than the interaction energy to rank possible pairs AB in the case of our synthetic data. The most efficient inference and ranking method here corresponds to using the Hungarian algorithm with the associated (global) energy gap score.

%\clearpage

\section{Impact of dataset size and quality on inference}
\label{sec_scaling}

After having shown that DCA can simultaneously solve the inference of couplings and of partners in our synthetic dataset, let us investigate in more quantitative detail the impact of dataset size and quality on these coupled inference problems.

\begin{figure}[h!]
	\begin{center}
		\includegraphics[keepaspectratio,width=\columnwidth]{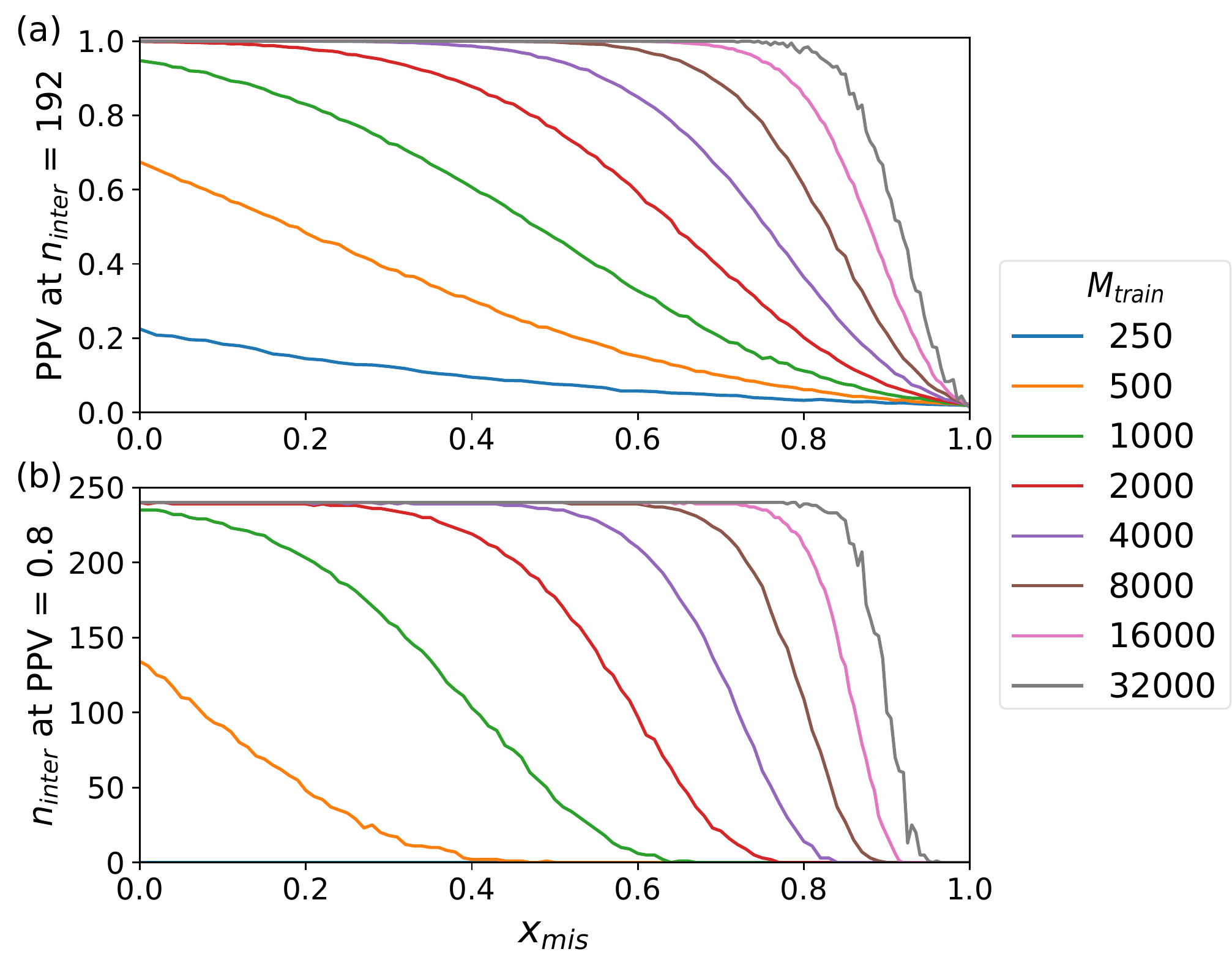}
	\end{center}
	\vspace{-5mm}
	\caption{(a) PPV for inter-block coupling prediction evaluated at the total number $n_{inter}=N_{inter}$ of actual inter-block couplings versus the fraction $x_{mis}=M_{mis}/M_{train}$ of mismatched chains AB in the training set. (b) Number $n_{inter}$ of correct inter-block coupling predictions obtained at PPV = 0.8  versus $x_{mis}$. Averages over 50 realizations corresponding to different training sets are shown. Data is generated using the same SBM(0.025,0.02) graph with $L = 100$ spins per single chain A or B as in Fig.~\ref{ppv_inter}, also at a sampling temperature $T = 5.0$.}\label{PPVinter_vs_xMis_N200}
\end{figure}

Fig. \ref{PPVinter_vs_xMis_N200}(a) shows the impact of the fraction $x_{mis}=M_{mis}/M_{train}$ of mismatched chains AB in the training set on the PPV for inter-block coupling prediction evaluated at the total number $n_{inter}$ of actual inter-block couplings for different training set sizes $M_{train}$. Fig. \ref{PPVinter_vs_xMis_N200}(b) shows the impact of $x_{mis}$ on the number of correct inter-block coupling predictions obtained at PPV = 0.8.  Both of these figures show that mismatches decrease the quality of predictions, consistently with Fig.~\ref{ppv_inter}(b). Furthermore, they demonstrate that the larger the training set, the more robust predictions become to mismatches (at equal mismatch fractions $x_{mis}$). In other words, for a large training set, good predictions can be obtained even if the fraction of mismatches is large. As a concrete example, for $M_{train}=32000$, the PPV at  $n_{inter}$ predictions is still very close to 1 (specifically, 0.98) for $x_{mis}=0.8$, which corresponds to a strongly corrupted dataset (see Fig. \ref{PPVinter_vs_xMis_N200}(a)).

\begin{figure}[h!]
	\begin{center}
		\includegraphics[keepaspectratio,width=\columnwidth]{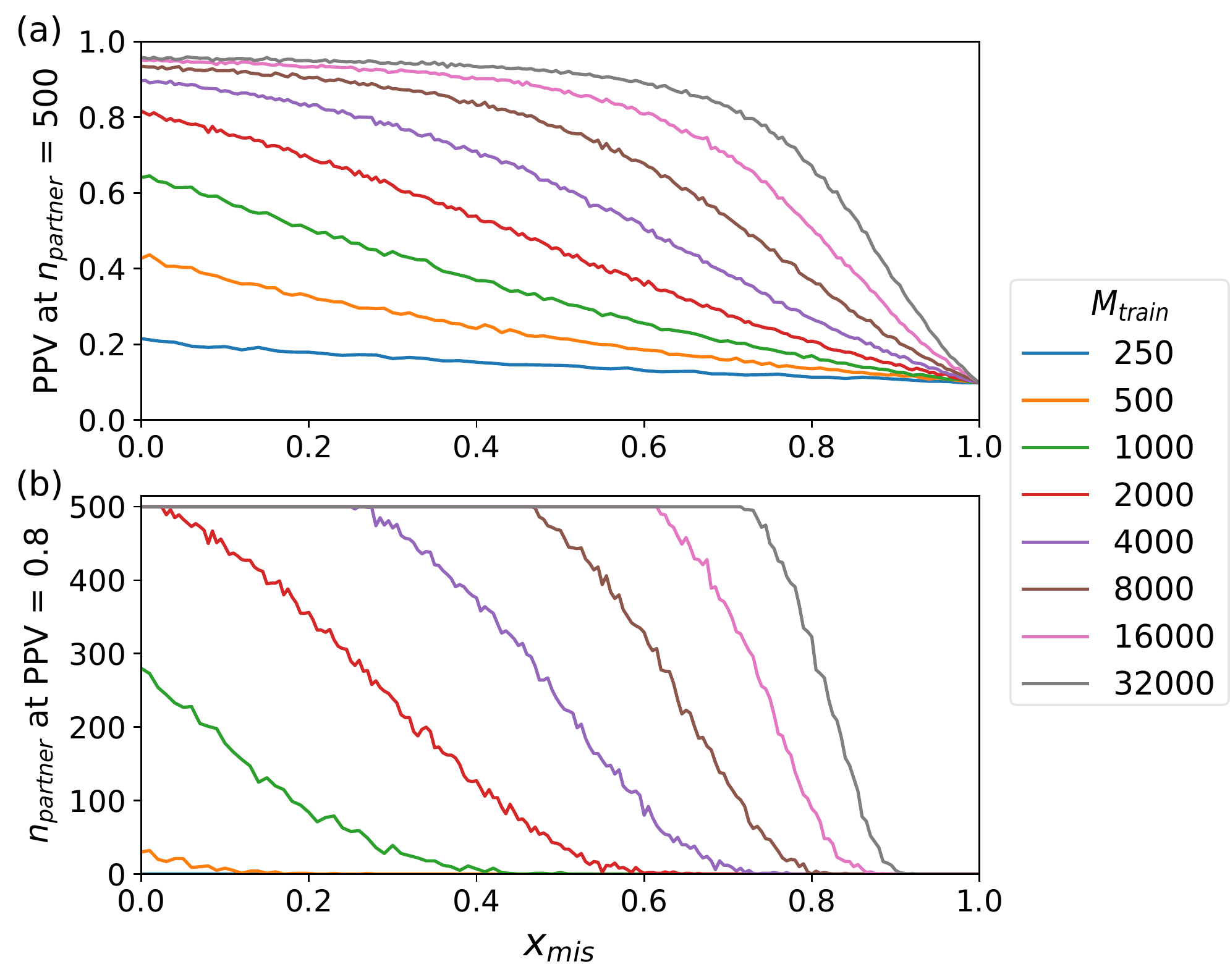}
	\end{center}
	\vspace{-5mm}
	\caption{(a) PPV for partner prediction evaluated at the total number of partners $n_{partner} = M_{test}$ of pairs of correct partners in the testing set versus the fraction $x_{mis}=M_{mis}/M_{train}$ of mismatched chains AB in the training set. (b) Number of correct partner predictions obtained at PPV = 0.8  versus $x_{mis}$. Partners were predicted employing the Hungarian algorithm and ranked by decreasing gap score. Data was generated using the same SBM(0.025,0.02) graph with $L = 100$ spins per single chain A or B as in Fig.~\ref{ppv_inter}, also at a sampling temperature $T = 5.0$. Training sets and testing sets  comprise $M_{train} = 2000$ and $M_{test} = 500$ paired chains AB respectively, and the testing sets were randomly partitioned into species with 10 pairs AB each. Averages over 50 realizations corresponding to different training and testing sets are shown.
}\label{PPVcontacts_vs_xMis_N200}
\end{figure}

Similarly, Fig. \ref{PPVcontacts_vs_xMis_N200}(a) shows the impact of the fraction $x_{mis}$ of mismatched chains AB in the training set on the PPV for partner prediction evaluated at the total number of partners $n_{partner} = M_{test}$ of pairs of correct partners in the testing set for different training set sizes $M_{train}$, and Fig. \ref{PPVcontacts_vs_xMis_N200}(b) shows the impact of  $x_{mis}$ on the number of correct partner predictions obtained at PPV = 0.8.  Overall, the same trends are observed as in Fig. \ref{PPVinter_vs_xMis_N200} for the inter-block coupling predictions, but the prediction of partners appears to be slightly more demanding in terms of training set size and quality than that of inter-block couplings. As a concrete example, for $M_{train}=32000$, the PPV at  $n_{partner} = M_{test}$ predictions is 0.67 for $x_{mis}=0.8$, significantly lower than the optimal value of 0.96 observed for $x_{mis}=0$ (see Fig. \ref{PPVcontacts_vs_xMis_N200}(a)), while the performance of inter-block coupling prediction is still almost optimal for such a training set (see Fig. \ref{PPVinter_vs_xMis_N200}(a)). Note however that the difficulty of the pairing task depends on the number of pairs per species, which we held fixed to 10 here.

\begin{figure}[h!]
	\begin{center}
		\includegraphics[keepaspectratio,width=\columnwidth]{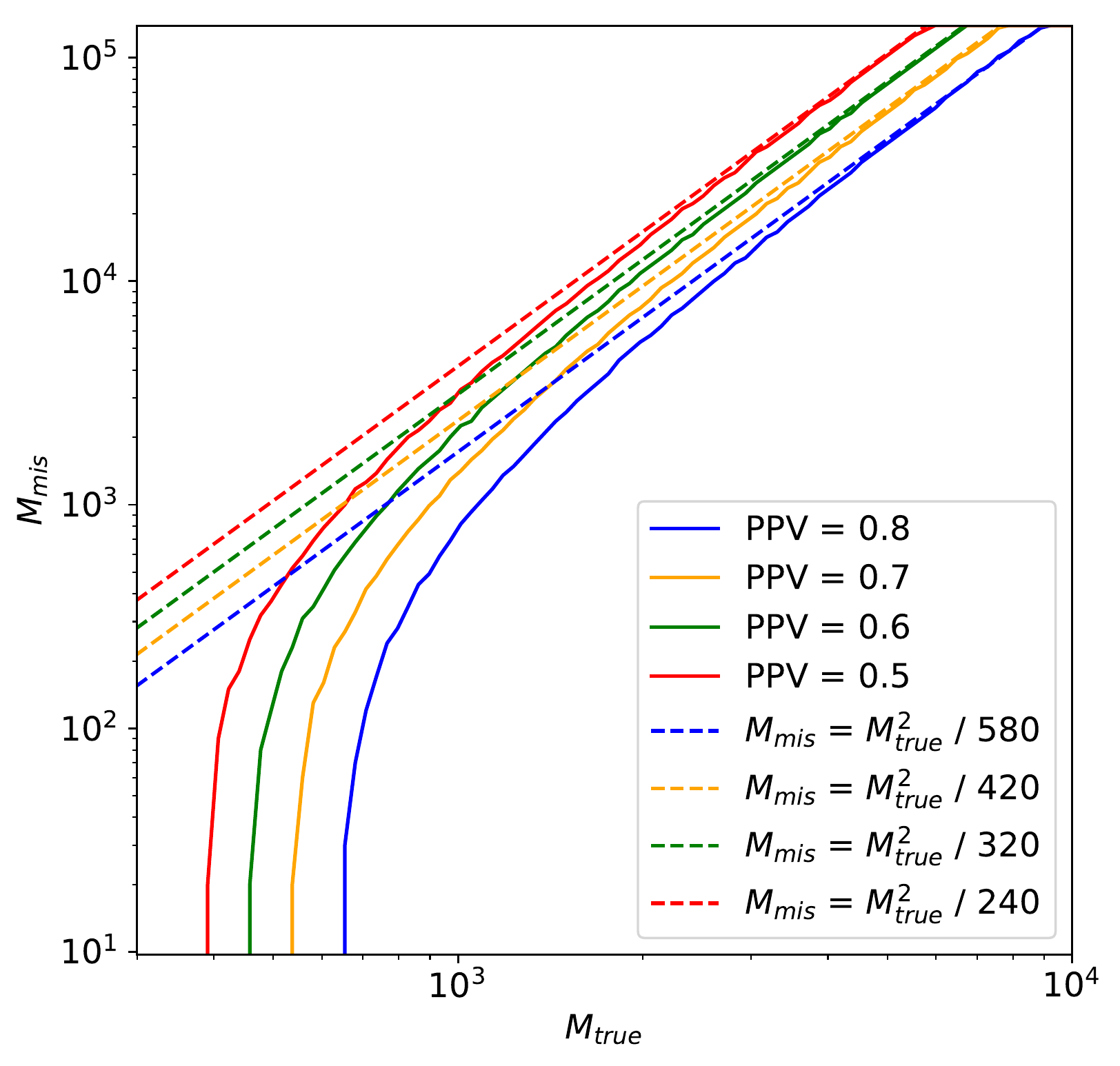}
	\end{center}
	\vspace{-5mm}
	\caption{Maximum number $M_{mis}$ of mismatched chain pairs AB that can be tolerated in a training set comprising $M_{true}$ correct pairs while still attaining a given average PPV value for inter-block coupling prediction evaluated at the total number $n_{inter}=N_{inter}$ of actual inter-block couplings. Data was generated using the same SBM(0.025,0.02) graph with $L = 100$ spins per single chain A or B as in Fig.~\ref{ppv_inter}, also at a sampling temperature $T = 5.0$. Averages over 100 realizations of the training set are shown (see App.~\ref{scaling_app}).
	}\label{inter_Mtrain_vs_Mmis_N200}
\end{figure}

\begin{figure}[h!]
	\begin{center}
		\includegraphics[keepaspectratio,width=\columnwidth]{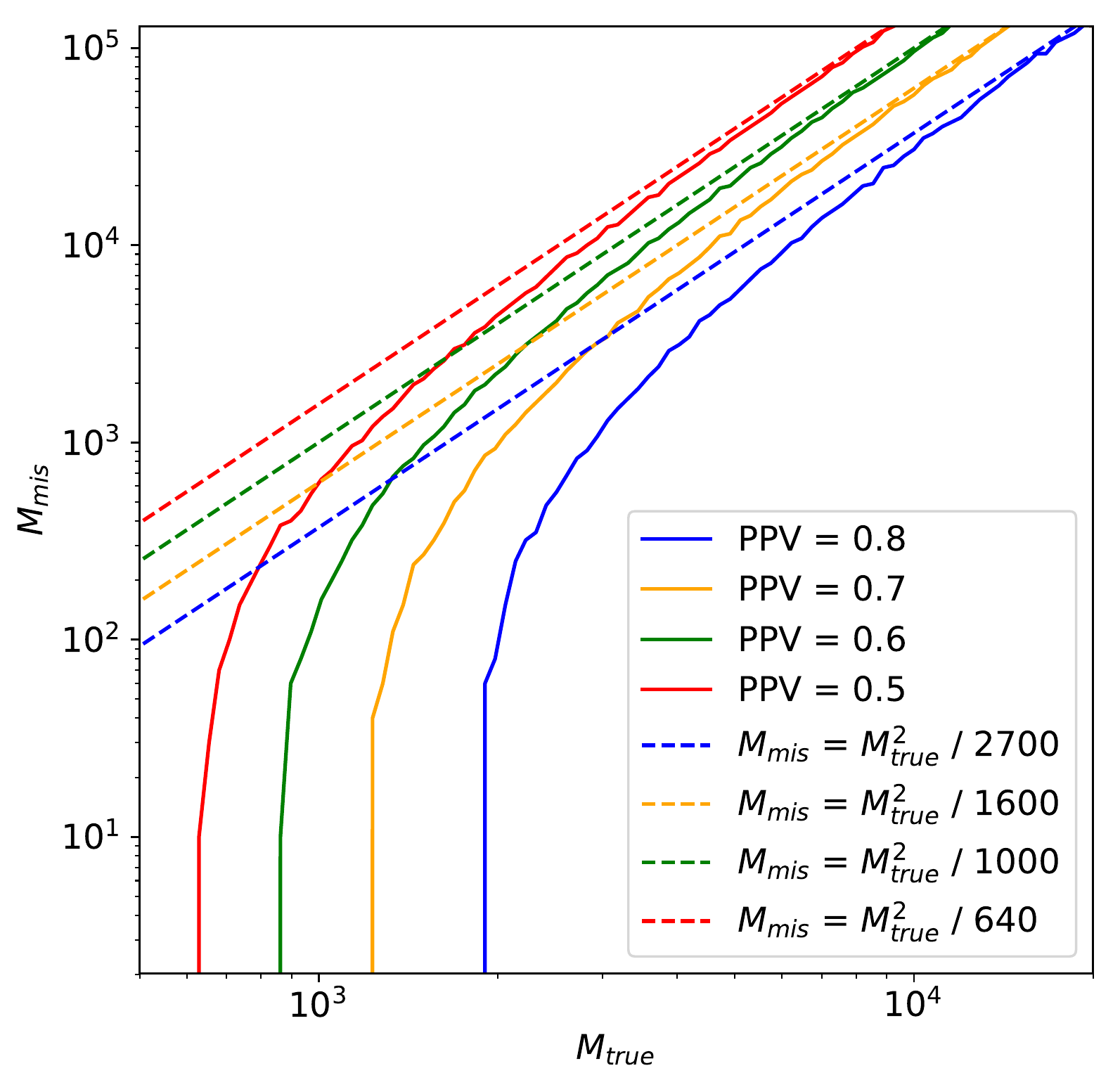}
	\end{center}
	\vspace{-5mm}
	\caption{Maximum number $M_{mis}$ of mismatched chain pairs AB that can be tolerated in a training set comprising $M_{true}$ correct pairs while still attaining a given PPV value for partner prediction evaluated at the total number of partners $n_{partner} = M_{test}$ of pairs of correct partners in the testing set. Data was generated using the same SBM(0.025,0.02) graph with $L = 100$ spins per single chain A or B as in Fig.~\ref{ppv_inter}, also at a sampling temperature $T = 5.0$. Testing sets comprise $M_{test} = 500$ paired chains AB respectively, and were randomly partitioned into species with 10 pairs AB each. Averages over 100 realizations of the training and testing sets are shown. 
		}\label{partner_Mtrain_vs_Mmis_N200}
\end{figure}

Let us now quantify the important observation that the larger the training set, the more robust both types of predictions become to mismatches (see Figs.~\ref{PPVinter_vs_xMis_N200} and~\ref{PPVcontacts_vs_xMis_N200}). In order to do this, we vary both the size of the training set $M_{train}$ and its quality via the number of mismatched pairs $M_{mis}$ in it. Specifically, for each number $M_{true}$ of correct pairs in the training set, we find the maximum number $M_{mis}$ of mismatched pairs that can be added to them (thus forming a training set of $M_{train}=M_{mis}+M_{true}$ pairs total) while still attaining a given inference performance, namely a given PPV value, either for inter-block contact predictions or for pairing prediction in a testing set. PPVs are averaged over many realizations of the training set and testing sets, see App.~\ref{scaling_app} for details of the procedure.
Figs.~\ref{inter_Mtrain_vs_Mmis_N200} and \ref{partner_Mtrain_vs_Mmis_N200} show that for both inference problems, with large enough training sets, the results scale like $M_{mis}\propto M_{true}^2$. This is consistent with the hypothesis that signal adds linearly, $\propto M_{true}$, while noise adds in a square-root fashion, $\propto \sqrt{M_{mis}}$, and that what matters for performance is the signal-to-noise ratio $\propto M_{true}/\sqrt{M_{mis}}$. This scaling quantifies our finding that lower quality can be tolerated for larger training sets.

The figures also show that in both problems a minimum number $M_{true}$ of correct matches in the training set is required to actually reach the desired PPV value. The almost vertical initial slope of the curves on Figs.~\ref{inter_Mtrain_vs_Mmis_N200} and \ref{partner_Mtrain_vs_Mmis_N200} shows the immediate robustness of the predictions as soon as this minimal value of $M_{true}$ has been crossed. The comparison of both the minimal $M_{true}$ and the allowed $M_{mis}$ at given $M_{true}$ illustrate again the fact that partner prediction is harder than coupling prediction for species comprising $k=10$ pairs AB (recall that the difficulty of partner prediction increases with $k$).

\newpage

\section{Making predictions without a training set}
\label{sec_noTS}

In Ref.~\cite{Bitbol16}, an Iterative Pairing Algorithm (IPA) based on DCA (DCA-IPA) was introduced to solve the partner prediction problem for two protein families, even in the absence of any training set. To initialize the iterative process without a training set, each protein chain A is randomly matched with a protein chain B from the same species. A DCA model is inferred from the resulting joint MSA of $M$ sequence pairs (for simplicity, the overall sequence number $M$ is assumed to be the same for both protein families). Next, the resulting interaction energies $E_{int}$ are used to predict new pairings between chains A and B, but only the $N_{increment}$ predictions with the largest energy gaps $\Delta E_{int}$ are used in order to learn a second DCA model and to make new partner predictions. The procedure is iterated, employing the first $(n-1)N_{increment}$ pairings in terms of $\Delta E_{int}$ from iteration $n-1$ to build the $n$th DCA model. Therefore, a growing number of paired chains is employed to build the DCA model at each iteration. The IPA terminates when $(n-1)N_{increment}=M$. 

\begin{figure}[h!]
	\begin{center}
		\includegraphics[keepaspectratio,width=0.9\columnwidth]{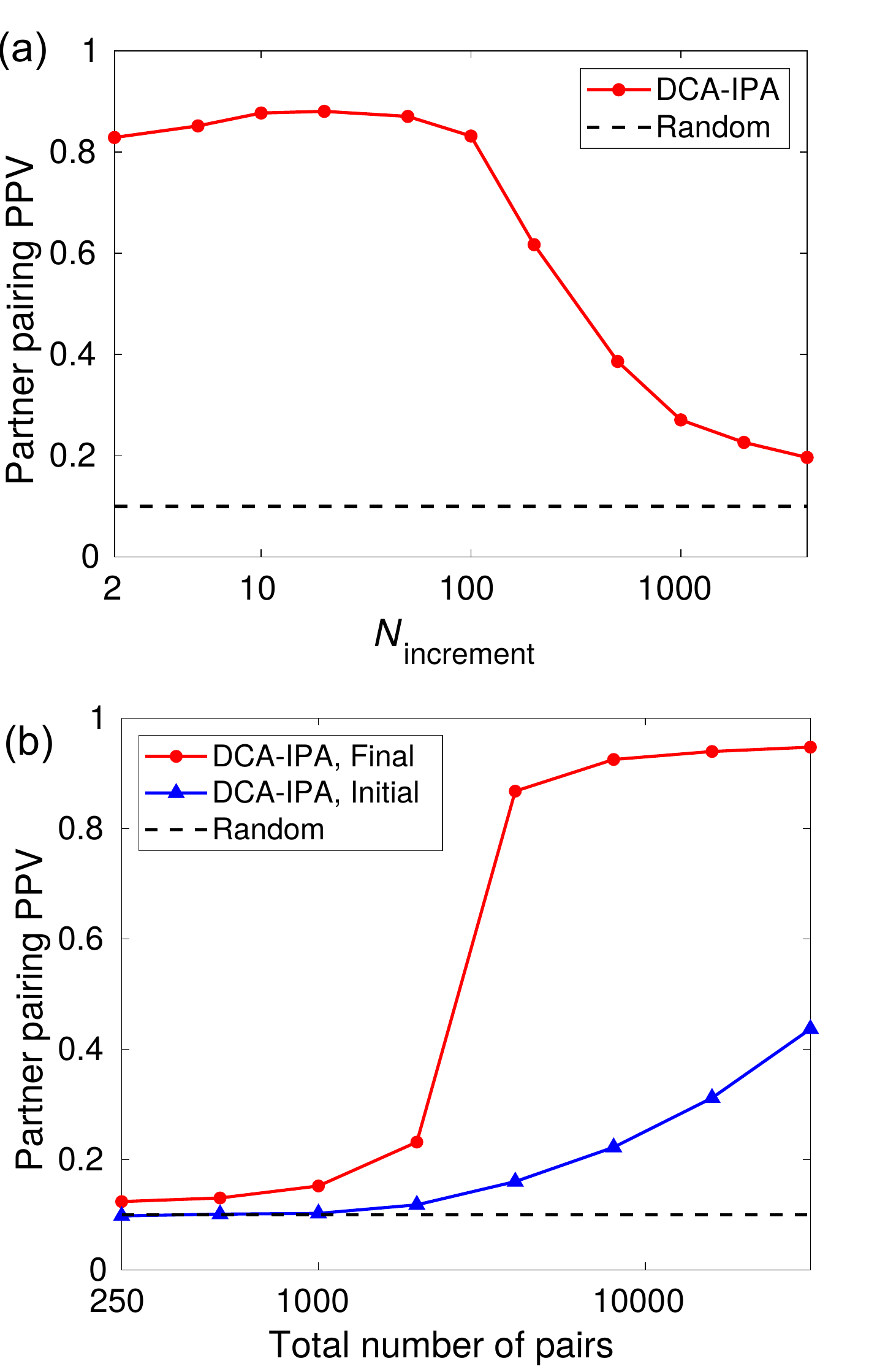}
	\end{center}
	\vspace{-5mm}
	\caption{(a) Positive predictive value (PPV) of pairing prediction, i.e. fraction of the pairs of chains AB that are correctly predicted by the DCA-IPA starting from no training set, plotted versus the increment step $N_{increment}$. (b) PPV of pairing prediction obtained in the initial and final iterations of the DCA-IPA starting from no training set, plotted versus the total number of pairs in the dataset. In both panels, data are generated using the same SBM(0.025,0.02) graph with $L = 100$ spins per single chain A or B as in Fig.~\ref{ppv_inter}, also at a sampling temperature $T = 5.0$. We then randomly extract (without replacement) a dataset from the total generated dataset, comprising a fixed number of complete chains AB ($4000$ in (a)). The chains in this dataset are randomly partitioned into species with 10 pairs AB each, and within each species, pairings are blinded. Averages over 50 realizations corresponding to different datasets are shown. The dashed line indicates the random expectation, corresponding to random within-species one-to-one matches.
		}\label{IPA1}
\end{figure}

Fig.~\ref{IPA1}(a) shows that for sufficiently small increment steps $N_{increment}$, the DCA-IPA allows us to make highly accurate pairing predictions for our controlled synthetic dataset even when starting without any training set. Furthermore, in order to obtain such a good performance, a large enough total dataset (large $M$) is required, as shown in Fig.~\ref{IPA1}(b).

Actually, the PPV of the first partner prediction based on random within-species pairings (blue curve) is very close to the one obtained from a dataset with a mismatch fraction  $x_{mis}=(k-1)/k=0.9$. Indeed, a random matching leads, on average, to one correct pair AB per species (i.e.~10\% in the case of $k=10$ sequences per species), and to $k-1$ mismatches. Given the results of the last section, the signal from the correct pairs adds more constructively than the noise from mismatches: this favors correct partners over mismatches, all the more that the dataset is large. Consequently, pairing performance results significantly above random expectation for large datasets, even when using the first DCA model.

The iterative process yields very large additional improvements (red curve), in particular for large datasets. We observe in Fig.~\ref{IPA1}(b) that the increase of PPV with dataset size is more abrupt with iterations than without, in good agreement with the sharp onset of the curves in Fig.~\ref{partner_Mtrain_vs_Mmis_N200}, which indicated a minimum training set size required for accurate partner prediction, but also an immediate robustness to noise.

\begin{figure}[h!]
	\begin{center}
		\includegraphics[keepaspectratio,width=\columnwidth]{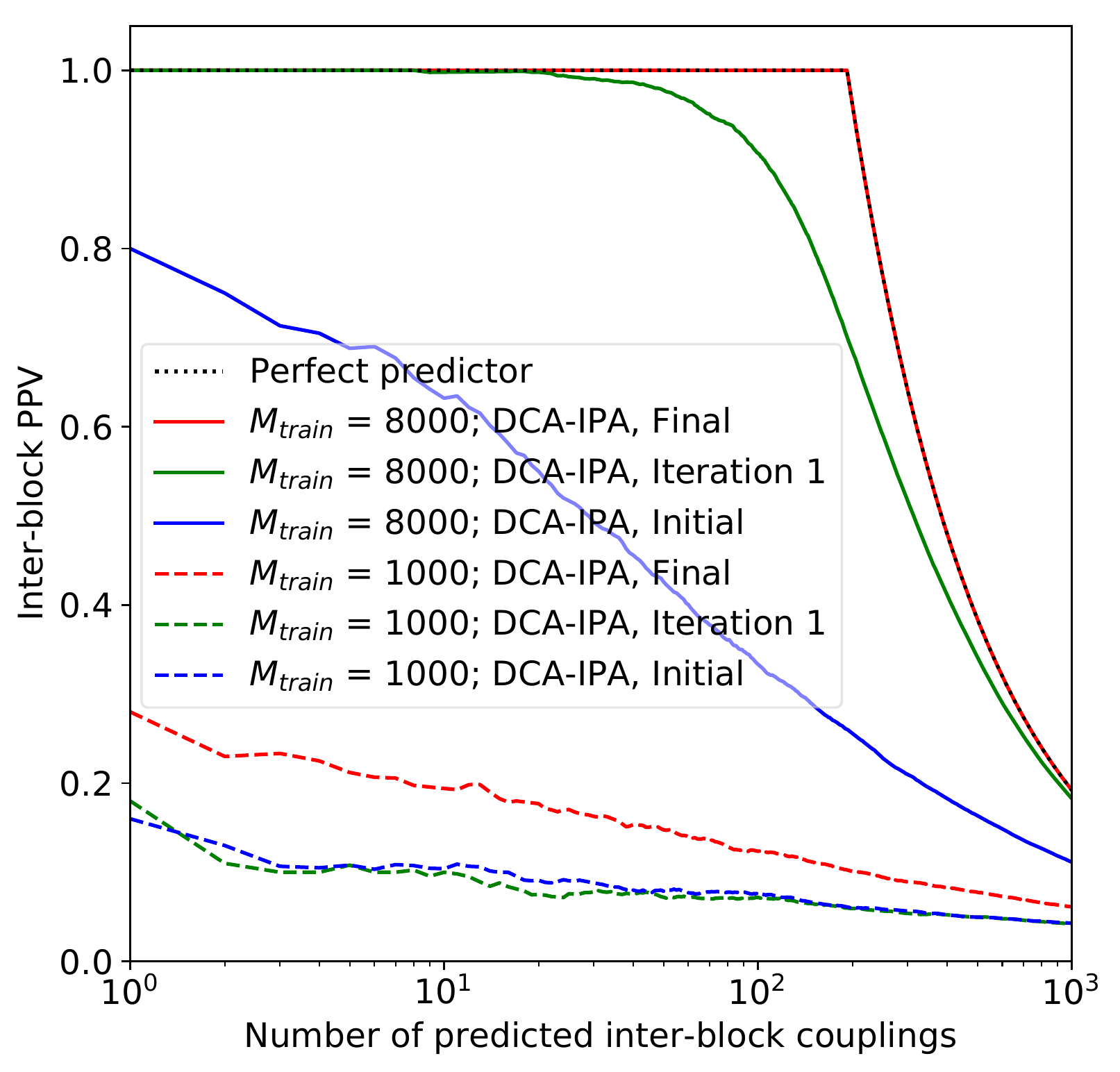}
	\end{center}
	\vspace{-5mm}
	\caption{Positive predictive value (PPV) of the inferred inter-block couplings, i.e. fraction of the top inferred inter-block couplings that are actual inter-block couplings in the stochastic block model (SBM) random graph used to generate the data, plotted versus the number of top inter-block couplings considered. Results are shown for two dataset sizes (1000 and 8000 paired chains AB), at the initial stage, i.e. for a DCA model constructed on a dataset of  random within-species one-to-one matches, after the first iteration, and after the final one.  Data was generated using the same SBM(0.025,0.02) graph with $L = 100$ spins per single chain A or B as in Fig.~\ref{ppv_inter}, also at a sampling temperature $T = 5.0$. We then randomly extract (without replacement) a training set from the total generated dataset, and DCA inference is performed. Averages over 50 realizations corresponding to different datasets are shown. }\label{IPA2}
\end{figure}

The IPA also allows to strongly improve the prediction of inter-block couplings, as shown in Fig.~\ref{IPA2}. Note however that for large enough datasets, even predictions from random matches are already much better than random, in agreement with the stochastic matching strategy of Ref.~\cite{Malinverni17}.

\clearpage

\section{Conclusion}
\label{Sec_ccl}
Conceptually, determining pairs of residues that are in contact in the three-dimensional structure of protein complexes and identifying interaction partners among paralogous proteins from sequence data are two coupled inference problems, which can be addressed together by Direct Coupling Analysis (DCA). In this paper, we have employed controlled synthetic data generated by Monte Carlo sampling of spin chains based on a stochastic block model Hamiltonian where the two blocks represent the two interacting proteins. This enabled us to assess the impact of dataset size and quality on their performance. Consistently with results obtained with natural data, we found that, provided that the training set contains $\sim\!1000$ synthetic sequences at least, DCA accurately identifies inter-chain and intra-chain couplings, as well as the pairs of synthetic sequences (spin chains) that actually come from one single chain generated by the model, and that model interaction partners in proteins. Furthermore, we show that mismatches between partners in the training set deteriorates the inference of inter-chain couplings and of other partners in a testing set, while being almost immaterial to the inference of intra-chain couplings. Both for inference of inter-chain couplings and for inference of partners, we found a quadratic scaling relating dataset quality and size that is consistent with noise adding in square-root fashion and signal adding linearly as dataset size is increased. This implies that it is often good to incorporate more data even if its quality is imperfect. An iterative pairing algorithm (IPA), where the top-scored predictions are gradually incorporated in the dataset employed to construct the DCA model, is useful to further increase performance, and allows to obtain good performance even in the absence of a training set.

It is important to note that in our synthetic data, all correlations are from actual couplings, excluding any impact of phylogeny. This is because our aim here is to quantitatively understand the bases of prediction of interaction partners thanks to contacts. By contrast, correlations arising in protein sequences due to their common evolutionary history, i.e. phylogeny~\cite{Casari95,Halabi09,Qin18,Fryxell96,Goh00,Pazos01,Ochoa10}, can contribute to the success of DCA-based approaches at predicting protein-protein interactions from real protein sequences~\cite{BitbolPMI,Marmier19}, while they obscure the identification of contacts~\cite{Weigt09,Marks11,Qin18}, thus leading to a more complex situation. Here, thanks to our controlled synthetic data sampled at equilibrium using Monte Carlo simulations, we isolated correlations from interactions from such other signals. Incorporating phylogeny, as well as considering Potts spins, including local fields, and incorporating the heterogeneity of interactions~\cite{Franco19} would all be interesting further directions making the synthetic data more realistic. This could give insight into what matters most for performance in natural protein sequence data.

Our results help to understand the success of inter-protein contact determination by DCA employing random within-species pairings between the paralogs of two protein families and considering the contacts appearing repeatedly in multiple random pairings~\cite{Malinverni17}. Indeed, remarkably, these very good results were obtained despite the presence of multiple incorrect pairings, but the dataset was large ($\sim20000$ sequences total in both protein families considered). Our results also shed light on the reasons of the success of iterative DCA-based methods at predicting interaction partners among the paralogs of ubiquitous prokaryotic protein families~\cite{Bitbol16,Gueudre16}, in particular on the fact that predictions can be made without any training set, starting from random within-species pairings~\cite{Bitbol16}. Indeed, at the first iteration, where random one-to-one pairings are made within each species, the expectation of the fraction of correct pairs is small, specifically $1/\langle k\rangle$ where $\langle k\rangle = M/N_{species}$ is the average number of sequences per species, and $N_{species}$ the number of species while $M$ is the total number of sequences. Nevertheless, predictions become better and better during the iterative process. The fact that signal from couplings adds more constructively than noise, as quantitatively demonstrated here, is one of the ingredients that explain this bootstrapping of the iterative pairing algorithm. Another one is that among pairs AB constructed by pairing a chain A and a chain B from the same species, correct pairs of partners possess more neighbors in terms of sequence similarity than incorrect pairs. Ref.~\cite{Bitbol16} called this the \textit{Anna Karenina effect}, in reference to the first sentence of Tolstoy's novel. This effect, which is strong in natural datasets~\cite{Bitbol16} as well as in datasets comprising phylogenetic correlations~\cite{Marmier19}, favors correct pairs in the iterative pairing algorithm, especially at early iterations~\cite{Bitbol16}. In natural data, we expect both of these effects to contribute to yield good performance in predicting interaction partners when starting without a training set.

\appendix

%\clearpage
\section{Data generation}
\label{App_data_gen}

We consider chains of spins generated from a Hamiltonian with pairwise couplings between sites linked by edges in a random graph. Therefore, the coupled sites represent the amino acids that are in contact in the case of proteins. For simplicity, we assume that all coupling strengths are the same, and set them to 1, thereby setting our energy unit. The associated Hamiltonian reads:
\begin{equation}
H(\vec{\sigma})=-\sum_{(i,j)\in \mathcal{E}}\sigma_i\sigma_j\,,
\label{ham}
\end{equation}
where $\vec{\sigma}=(\sigma_1,\dots,\sigma_{2L})$ is the chain of spins with total length $2L$ and $\mathcal{E}$ is the set of edges of the random graph considered.

For our random graph, we consider a stochastic block model~\cite{dyer1989solution}, a well-known random graph model, which is used in statistics, machine learning and network science to represent modular networks. This model incorporates blocks with different connectivities inside the blocks and across them. Here, we consider two blocks A and B, each of them representing a protein, so that intra-block couplings represent intra-protein contacts while inter-block couplings represent inter-protein contacts. This allows us to incorporate the fact that inter-protein contacts tend to be sparser than intra-protein ones. In practice, each of our two blocks comprises $L=100$ vertices, yielding 200 vertices total, and edges between pairs of vertices are chosen randomly among all possible pairs, with fixed probabilities $p_{intra}=0.025$ inside the blocks and $p_{inter}=0.02$ across them. We construct the random graph once, and retain it throughout. The chosen graph has 192 inter-block edges and 260 intra-block edges. In order to assess the possible impact of finite size effects, we also consider a graph with $L=50$ vertices in each block and probabilities $p_{intra}=0.05$ inside the blocks and $p_{inter}=0.04$ across them, so that the number of edges per vertex is the same on average in our two graphs, ensuring similar properties such as the phase transition temperature.

\begin{figure}[htb]
	\begin{center}
		\includegraphics[keepaspectratio,width=\columnwidth]{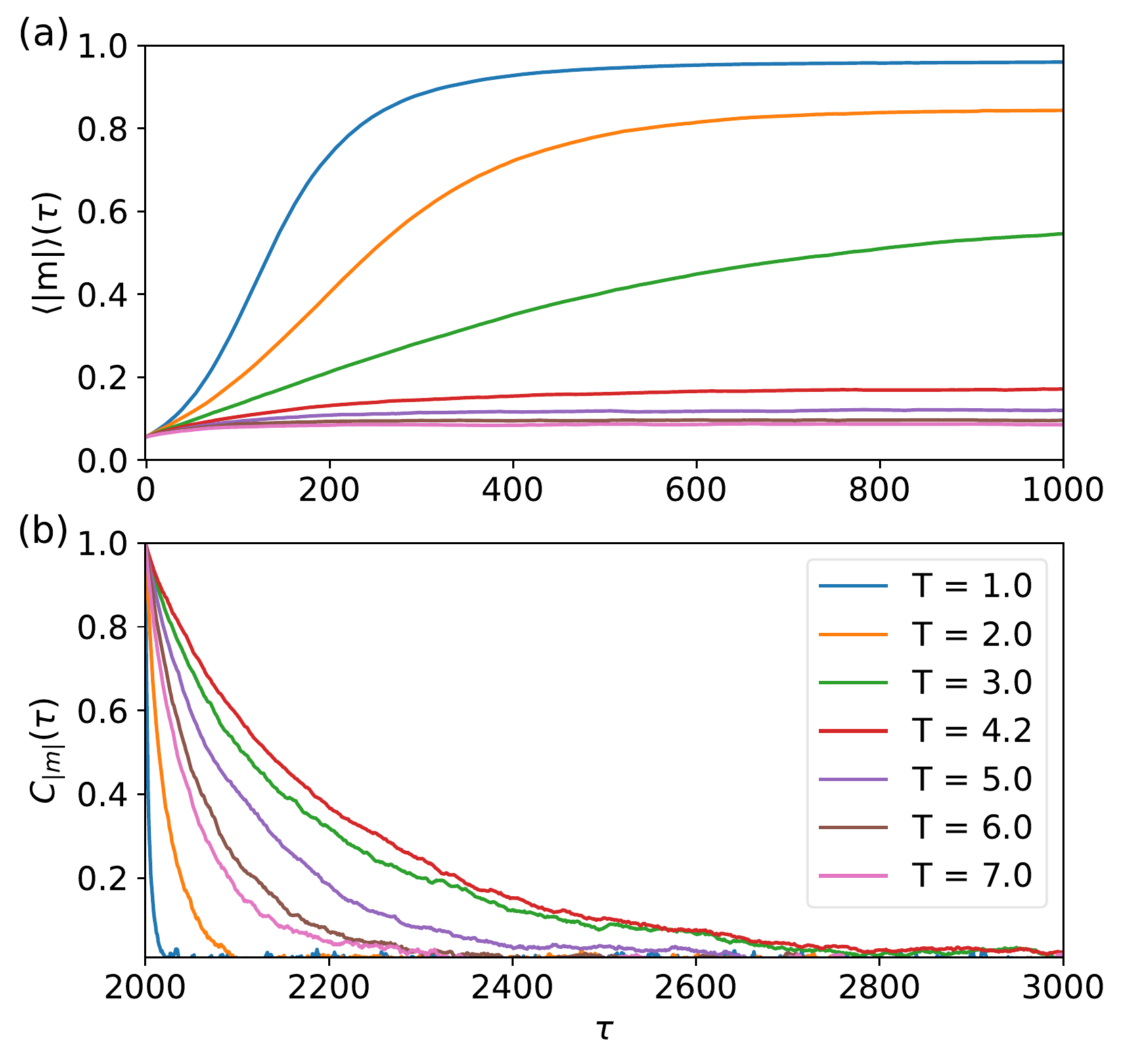}
	\end{center}
	\vspace{-5mm}
	\caption{(a) Average absolute value of magnetization $\langle|m|\rangle$ plotted versus the number $\tau$ of accepted spin flips. (b) Auto-correlation of the absolute magnetization $|m|$ (see Eq.~\ref{cor}) plotted versus the number $\tau$ of accepted spin flips.  Data was generated using the same SBM(0.025,0.02) graph with $L = 100$ spins per single chain A or B as in Fig.~\ref{ppv_inter}, and different sampling temperatures $T$ were employed. Statistical averages denoted by $\langle .\rangle$ were computed over 10000 spin chains. }\label{autocorrelation_fig_v2}
\end{figure}

Data is generated employing a Monte Carlo sampling procedure according to the Hamiltonian in Eq.~\ref{ham} with the set of edges $\mathcal{E}$ defined by our stochastic block model random graph. Specifically, for each chain of spins, we start from a random chain of spins, propose spin flips, and accept them or not using the Metropolis criterion, i.e. all flips that lower the energy computed according to Eq.~\ref{ham} are accepted, while those that raise it are accepted with a probability $e^{-\Delta H/T}$ where $\Delta H$ is the energy variation associated to the spin flip, while $T$ is the sampling temperature, and we have set the Boltzmann constant to 1. We continue flipping spins until 2000 spin flips have been accepted, which is sufficient for the system to reach equilibrium. Indeed, the average of the absolute value of the magnetization of the chain
\begin{equation}
m=\sum_{i=1}^{2L}\sigma_i
\label{mag}
\end{equation}
converges after $\sim\!1000$ accepted spin flips for all sampling temperatures considered, as shown in Fig.~\ref{autocorrelation_fig_v2}(a). Furthermore, $\sim\!1000$ accepted spin flips are also enough to lose correlation between an equilibrated spin chain and a further evolved one, as shown in Fig.~\ref{autocorrelation_fig_v2}(b). Specifically, the correlation of absolute magnetization is shown versus the number of accepted spin flips $\tau$ in Fig.~\ref{autocorrelation_fig_v2}(b):
\begin{equation}
C_{|m|}(\tau)=\frac{\langle|m(t)||m(t+\tau)|\rangle-\langle|m(t)|\rangle\langle|m(t+\tau)|\rangle}{\langle|m(t)|^2\rangle-\langle|m(t)|\rangle^2}\,,
\label{cor}
\end{equation}
where $t$ is chosen so that equilibrium has been reached, in practice $t=2000$. Note that therefore the correlation value does not depend on $t$. 

\begin{figure}[htb]
	\begin{center}
		\includegraphics[keepaspectratio,width=\columnwidth]{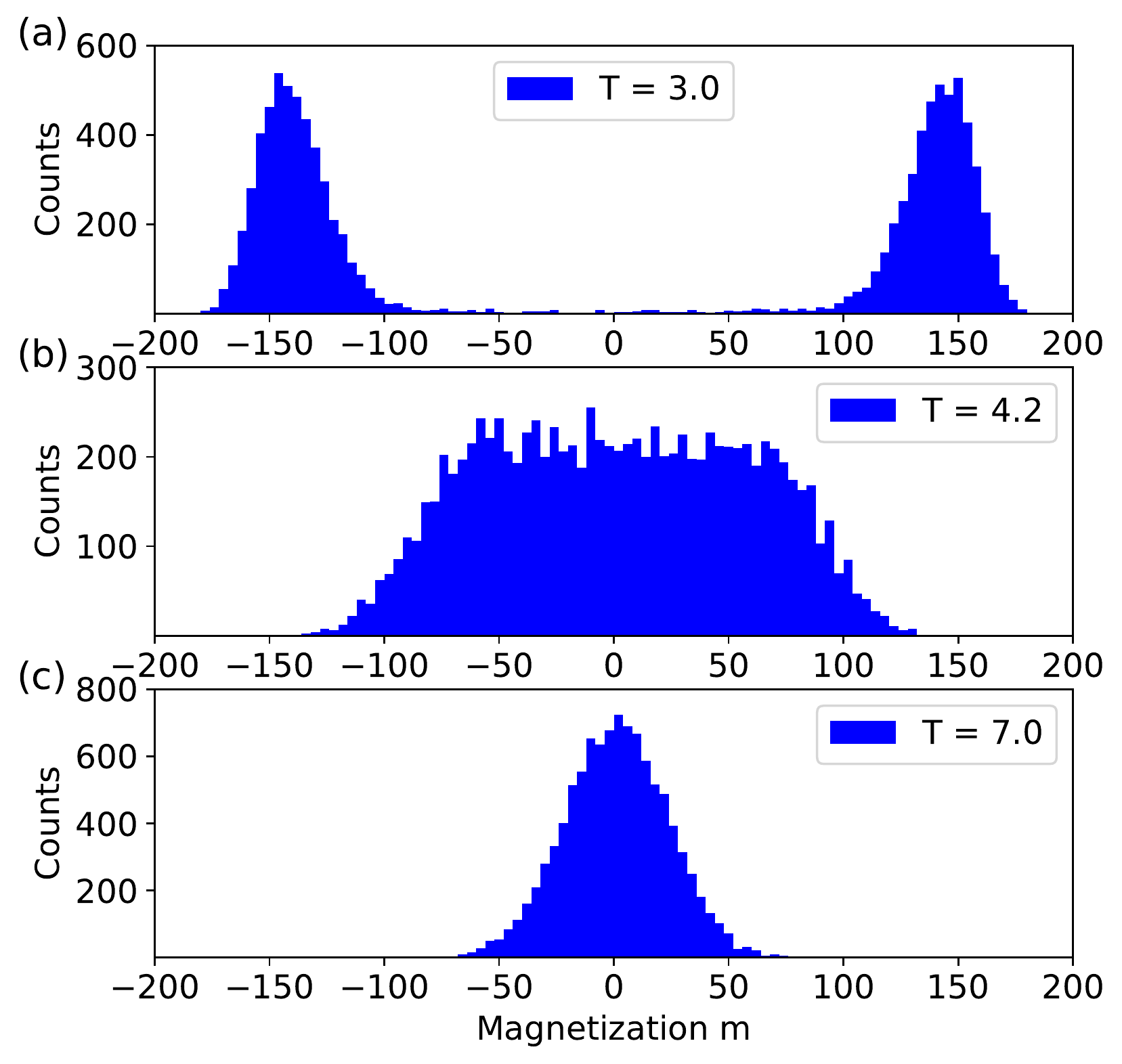}
	\end{center}
	\vspace{-7mm}
	\caption{Histograms of magnetization $m$ at different sampling temperatures. Data was generated using the same SBM(0.025,0.02) graph with $L = 100$ spins per single chain A or B as in Fig.~\ref{ppv_inter}. Each histogram was computed over 10000 spin chains.}\label{histog}
\end{figure}

An important parameter of the data generation process is the sampling temperature $T$. In particular we expect a ferromagnetic-paramagnetic transition as $T$ is increased. This can be seen on Fig.~\ref{autocorrelation_fig_v2}(a), as the absolute magnetization is close to zero for small temperature and close to one for large ones. Furthermore, Fig.~\ref{autocorrelation_fig_v2}(b) is indicative of a slowing down of the relaxation of correlation for $T$ close to 4.2, which hints at a phase transition at this temperature. The phase transition can be studied in more detail by plotting the histograms of magnetization. As shown in Fig. \ref{histog}, there is a switch between a bimodal distribution at large absolute magnetization, either negative or positive, for small $T$, corresponding to the ferromagnetic phase (Fig. \ref{histog}(a)), and a unimodal one centered on 0 for large $T$ (Fig. \ref{histog}(c)), corresponding to the paramagnetic phase. The value of $T$ where the histogram of magnetization becomes wide and flat corresponds to the phase transition temperature: here it is close to $T=4.2$ (Fig. \ref{histog}(b)), which matches well the slowing-down observation on Fig.~\ref{autocorrelation_fig_v2}(b).

\section{Inference method}
\label{App_infer}

In DCA~\cite{Weigt09,Morcos11,Marks11,Cocco18}, one starts from the empirical covariances measured between all pairs of sites $(i,j)$ in the training set: $C_{ij}=\langle\sigma_i\sigma_j\rangle-\langle\sigma_i\rangle\langle\sigma_j\rangle$, where $\sigma_i$ represents the spin state at site $i$ ($-1$ or $1$) and the average denoted by brackets ranges over all chains of the training set. Importantly, here, we are considering complete chains AB, and $i$ and $j$ range from 1 to the total length $2L$ of such a chain (where $L$ is the length of chain A or B). DCA is based on building a global statistical model from these covariances (and the one-body frequencies)~\cite{Weigt09,Morcos11,Marks11,Cocco18}, through the maximum entropy principle~\cite{Jaynes57}. This results in a $2L$-body probability distribution $P$ of observing a given sequence $(\sigma_1,\dots,\sigma_{2L})$ that reads $P(\sigma_1,\dots,\sigma_{2L})=\exp\left[\sum_{i<j}J_{ij}\sigma_i\sigma_j+\sum_{i=1}^{2L} h_i\sigma_i\right]/Z$, where $Z$ is the partition function and ensures normalization. This corresponds to the Boltzmann distribution associated to an Ising model with couplings $J_{ij}$ and fields $h_i$~\cite{Cocco18}. Inferring the coupling strengths that appropriately reproduce the empirical covariances is a difficult problem, known as the inverse Ising problem~\cite{Nguyen17}. Within the mean-field approximation, which is employed here, these coupling strengths can be approximated by $J_{ij}=-C^{-1}_{ij}$~\cite{Plefka82,Morcos11,Marks11}. 

The effective interaction energy $E_{int}$ of each possible pair AB in the testing set, constructed by concatenating a chain A and a chain B, can then be assessed via 
\begin{equation}
E_{int}=-\sum_{i=1}^{L}\sum_{j=L+1}^{2L} J_{ij}\sigma_i^A\sigma_j^B\,.
\label{energy}
\end{equation}
In real proteins, approximately minimizing such a score has proved successful at predicting interacting partners~\cite{Bitbol16}. Note that we only sum over inter-chain pairs (i.e. pairs of sites involving one site in A and one in B) because we are interested in interactions between A and B.

\section{Impact of sampling temperature and chain length on inference}
\label{app_length}

\begin{figure}[h!]
	\begin{center}
		\includegraphics[keepaspectratio,width=0.5\textwidth]{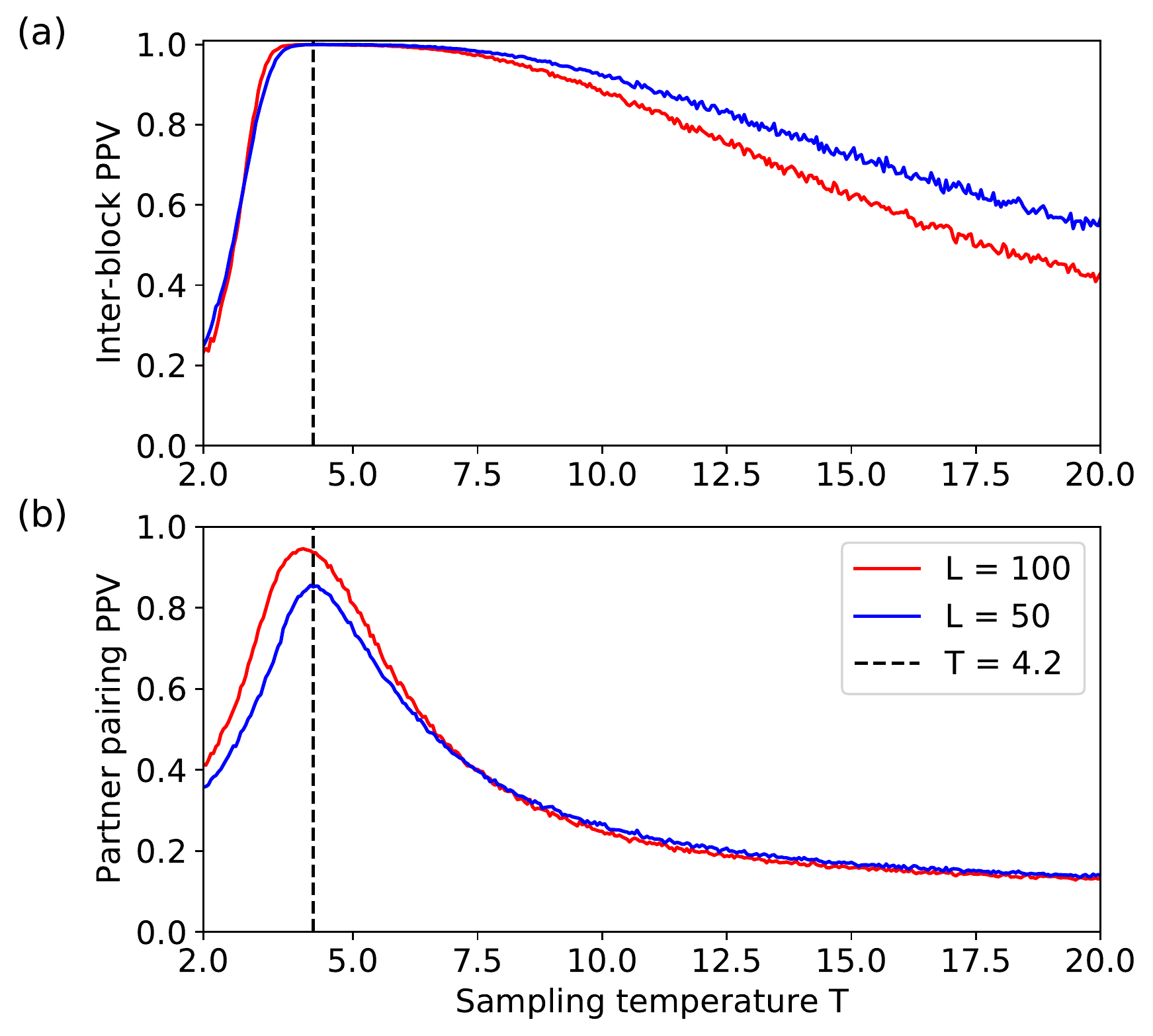}
	\end{center}
	\vspace{-5mm}
	\caption{(a) PPV for inter-block coupling prediction evaluated at the total number $n_{inter}=N_{inter}$ of actual inter-block couplings, plotted versus the sampling temperature $T$.  (b) PPV for partner prediction evaluated at the total number of partners $n_{partner} = M_{test}$ of pairs of correct partners in the testing set, plotted versus the sampling temperature $T$. Partners were predicted employing the Hungarian algorithm and ranked by decreasing gap score. In both panels, the vertical black dashed line indicates the approximate transition temperature $T=4.2$. In both panels, for $L = 100$, data were generated using the same SBM(0.025,0.02) graph as in Fig.~\ref{ppv_inter}. For $L = 50$, data were generated using an SBM(0.05,0.04) graph  (the same one in both panels). Training sets comprise $M_{train} = 2000$ paired chains AB, and in (b), testing sets include $M_{test} = 500$ paired chains AB, and were randomly partitioned into species with 10 pairs AB each. Averages over 100 realizations corresponding to different training (and different testing sets in (b)) are shown.}\label{PPVinter_vs_T_N100-200}
\end{figure}

Throughout this paper, we have employed a sampling temperature $T=5$ to generate data. This is slightly above the ferromagnetic-paramagnetic transition in our model, as can be seen on Figs.~\ref{autocorrelation_fig_v2} and Fig. \ref{histog}, which show that the phase transition temperature is close to $T=4.2$. It is interesting to investigate the impact of sampling temperature on the performance of inference. Fig.~\ref{PPVinter_vs_T_N100-200} demonstrates that maximal performance for the inference of inter-block couplings and of partners is obtained close to the phase transition temperature, and that the temperature range where performance is near-optimal is wider for inter-block coupling inference (Fig.~\ref{PPVinter_vs_T_N100-200}(a)) than for partner inference (Fig.~\ref{PPVinter_vs_T_N100-200}(b)). In addition, Fig.~\ref{PPVinter_vs_T_N100-200} shows the robustness of our results to changing the length of single chains A or B from $L=100$ to $L=50$ while simultaneously doubling the probabilities $p_{intra}$ and $p_{inter}$ of presence of edges used to generate the graph, so that the number of edges per vertex is similar in the two specific graphs considered.

\clearpage

\section{Predicting intra-chain couplings}
\label{app_intrachain}

\begin{figure}[h!]
	\begin{center}
		\includegraphics[keepaspectratio,width=\columnwidth]{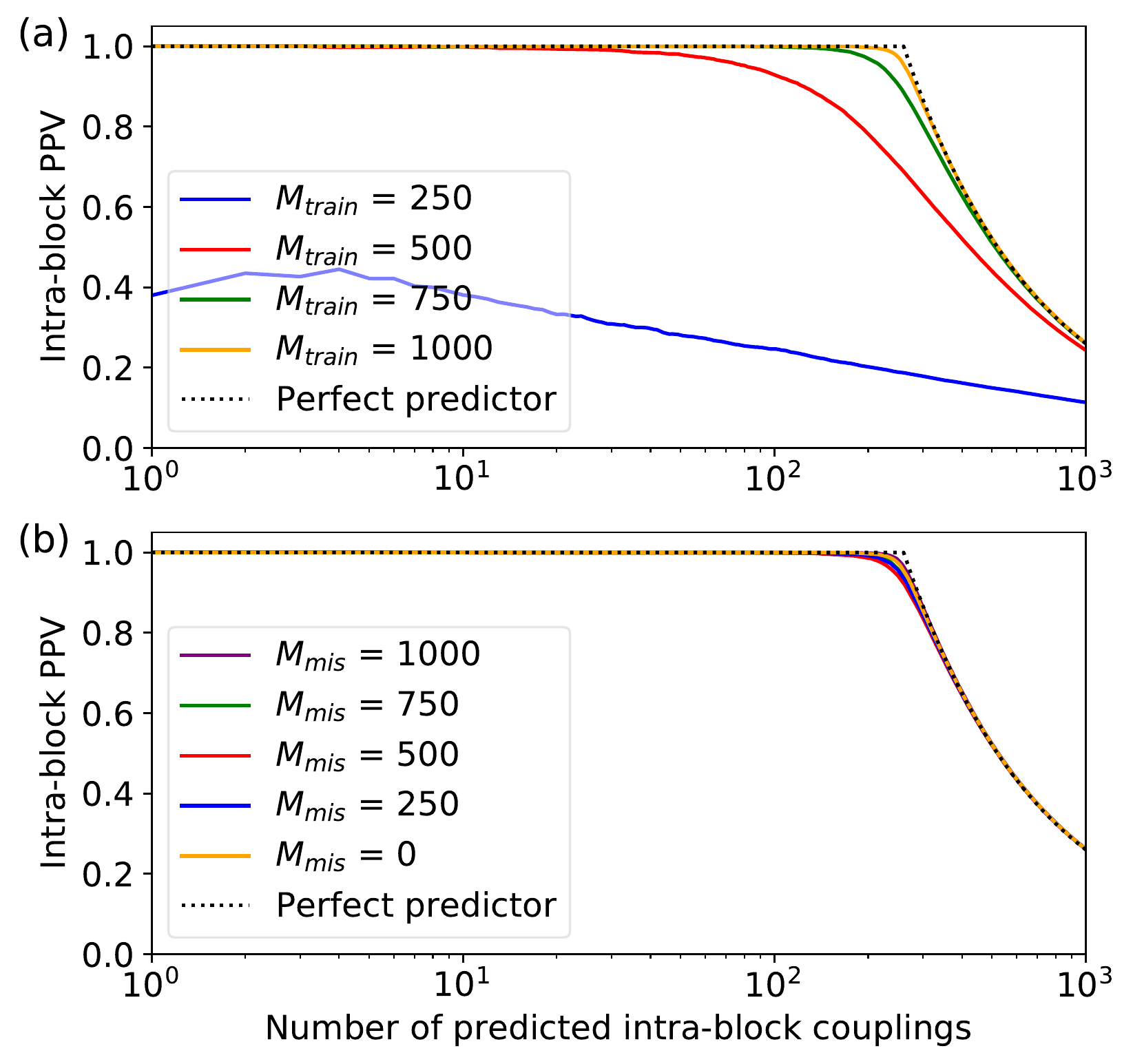}
		\end{center}
		\vspace{-5mm}
		\caption{PPV of the inferred intra-block couplings, i.e. fraction of the top inferred intra-block couplings that are actual intra-block couplings in the stochastic block model (SBM) random graph used to generate the data, plotted versus the number of top intra-block couplings considered. Data is generated using the same SBM(0.025,0.02) graph with $L = 100$ spins per single chain A or B as in Fig.~\ref{ppv_inter}, also at a sampling temperature $T = 5.0$. We then randomly extract (without replacement) a training set from the total generated dataset, and DCA inference is performed. Averages over 100 realizations corresponding to different training sets are shown. (a): Training sets with no mismatches comprising different numbers $M_{train}$ of chains AB. (b): Training sets with total number $M_{train}=1000$ of chains AB, comprising different numbers $M_{mis}$ of mismatched pairs AB. }\label{ppv_intra}
		\end{figure}

While Fig. \ref{ppv_inter} showed the impact of training set size and quality on the performance of inference of inter-block couplings, here we show in Fig. \ref{ppv_intra} the case of intra-block couplings. Fig. \ref{ppv_inter}(a), which is obtained in the absence of mismatched pairs in the training set, is extremely similar to Fig. \ref{ppv_intra}(a). Indeed, with perfectly paired chains AB in the training set, we do not expect inter- and intra-block couplings to behave differently. Furthermore, we observe no deterioration of the inference of intra-block couplings due to the mismatches in Fig. \ref{ppv_intra}(b), which stands in contrast with the case of inter-block couplings shown in Fig. \ref{ppv_inter}(b).

\clearpage

\section{Construction of the scaling plots}
\label{scaling_app}

\begin{figure}[htb]
	\begin{center}
		\includegraphics[keepaspectratio,width=\columnwidth]{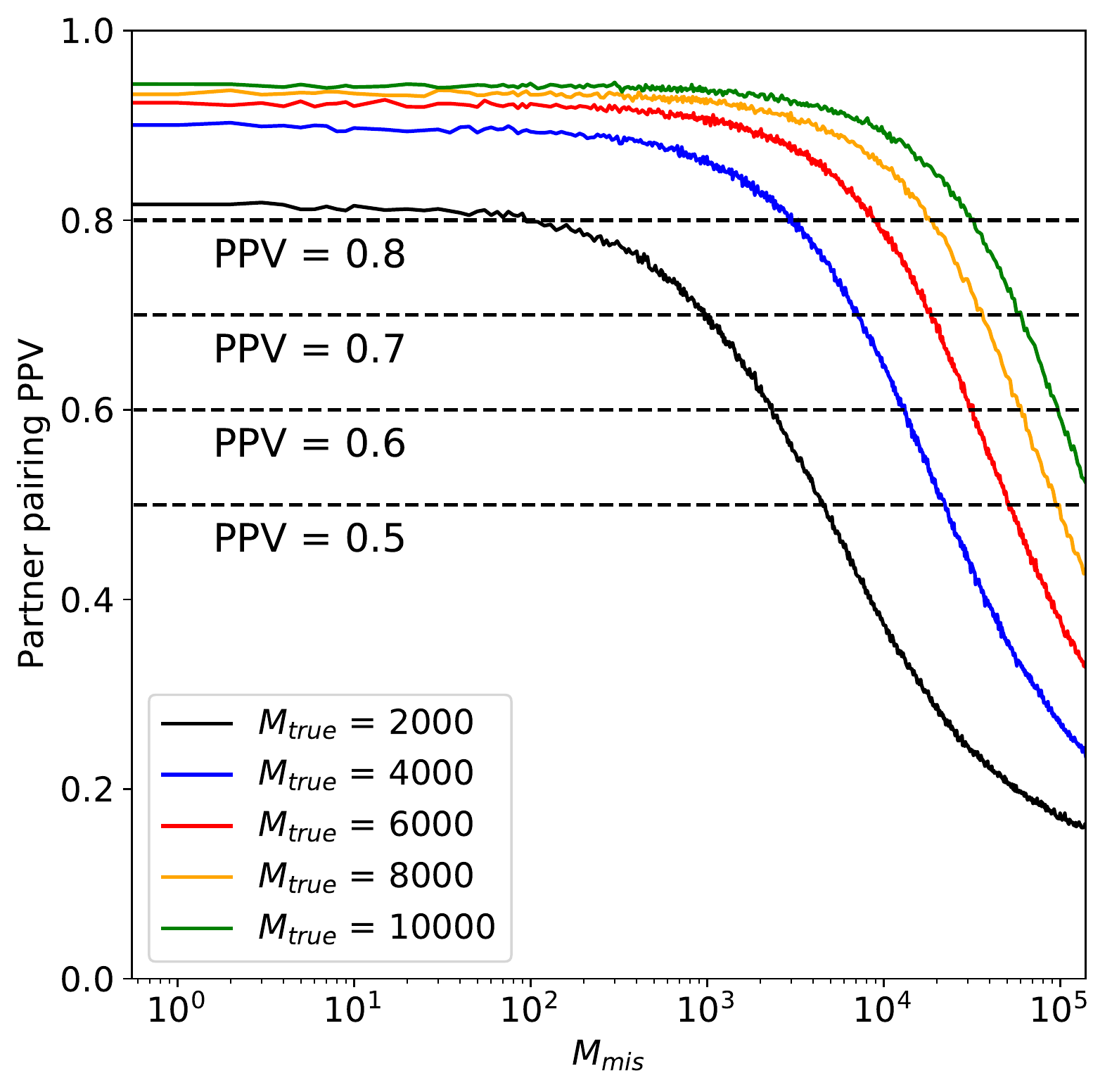}
	\end{center}
	\vspace{-5mm}
	\caption{PPV for partner prediction evaluated at the total number of partners $n_{partner} = M_{test}$ of pairs of correct partners in the testing set, plotted versus the number $M_{mis}$ of mismatched pairs AB in the training set, for various numbers $M_{true}$ of correct pairs in the training set. Data was generated using the same SBM(0.025,0.02) graph with $L = 100$ spins per single chain A or B as in Fig.~\ref{ppv_inter}, also at a sampling temperature $T = 5.0$.  Testing sets  comprise $M_{test} = 500$ paired chains AB respectively, and were randomly partitioned into species with 10 pairs AB each. Averages over 100 realizations are shown. In each realization, $M_{true}$ correct training pairs, $M_{mis}$ mismatched training pairs and $M_{test}$ testing pairs are randomly sampled from a large dataset of pairs.}\label{constr_scaling}
\end{figure}

In Figs.~\ref{inter_Mtrain_vs_Mmis_N200} and \ref{partner_Mtrain_vs_Mmis_N200}, we quantified the impact of dataset size and quality on our two coupled inference tasks. Here, we explain in detail how these plots were constructed. Specifically, for each number $M_{true}$ of correct pairs in the training set, we investigated how many mismatched pairs $M_{mis}$ can be added to the training set while still attaining a given inference performance, namely a given PPV, either for inter-block contact predictions (Fig.~\ref{inter_Mtrain_vs_Mmis_N200}) or for pairing prediction in a testing set (Fig.~\ref{partner_Mtrain_vs_Mmis_N200}).  To this end, for each given value of $M_{true}$, we gradually increased $M_{mis}$ and computed the corresponding PPV, averaged over many replicates, as shown in Fig.~\ref{constr_scaling}. Each replicate corresponds to a different realization of the training set (and for partner prediction, of the testing set), constructed by randomly picking different pairs from the large dataset of 150,000 spin chains AB constructed as explained in App.~\ref{App_data_gen}. For each value of $M_{true}$, the value of $M_{mis}$ such that a given PPV is reached can be read off from these results: in Fig.~\ref{constr_scaling}, it corresponds to the intersections between the colored curves and the dashed lines.

\clearpage

\section{Robustness to different graph connectivities}
\label{connectivities_app}

So far, we have mainly focused on one specific realization of an SBM graph having two blocks A and B, each one with $L=100$ vertices, and intra- and inter-block connectivities of $p_{intra}=0.025$ and $p_{inter}=0.02$ respectively, and the impact of $L$ at constant number of edges per vertex was discussed in App.~\ref{app_length}. In order to further test the robustness of our conclusions to the values of these connectivities, and motivated by the fact that inter-protein contacts are sparse, and often sparser than intra-protein ones, we now consider variants of this graph:
\begin{itemize} 
\item an SBM graph with the same parameters as our baseline model ($L=100$ and $p_{intra}=0.025$), except for a lower inter-block connectivity $p_{inter}=0.005$;
\item a graph called ``Interfaces 1'' where inter-block couplings are restricted to a random subset of sites of each block, namely 20 sites out of 100 in each of the two blocks, and with $p_{inter}=0.125$, thus featuring the same average number (50) of inter-block contacts as the first variant described above.
\item a graph called ``Interfaces 2'' that has the same intra-block contacts as SBM(0.025,0.02) and 57 inter-block contacts mainly located in two small subsets of sites of each of the two blocks.
\end{itemize}

\begin{figure}[h!]
	\begin{center}
		\includegraphics[keepaspectratio,width=0.5\textwidth]{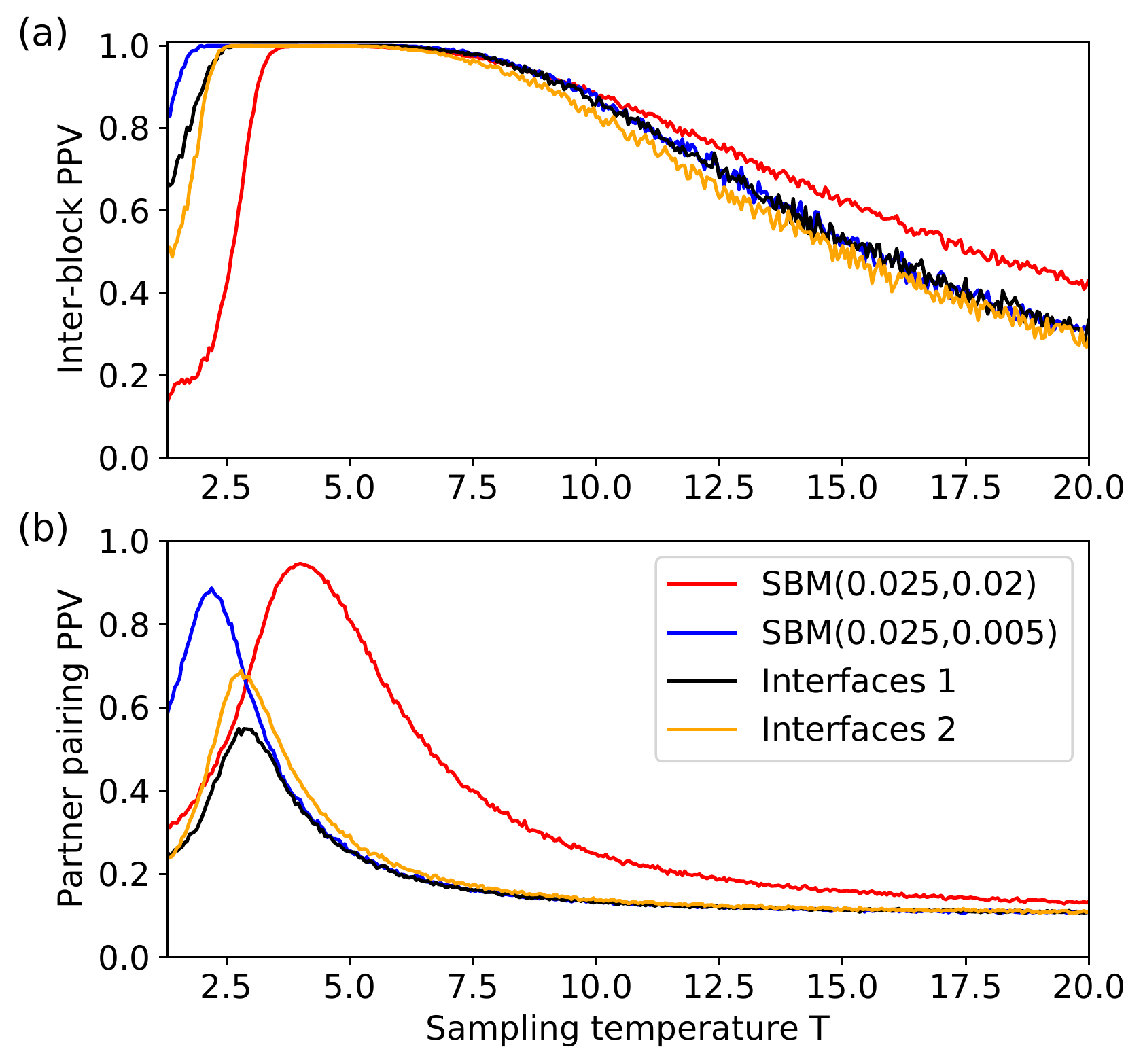}
	\end{center}
	\vspace{-5mm}
	\caption{(a) PPV for inter-block coupling prediction evaluated at the total number $n_{inter}=N_{inter}$ of actual inter-block couplings, plotted versus the sampling temperature $T$.  (b) PPV for partner prediction evaluated at the total number of partners $n_{partner} = M_{test}$ of pairs of correct partners in the testing set, plotted versus the sampling temperature $T$. Partners were predicted employing the Hungarian algorithm and ranked by decreasing gap score. Red curves correspond to data generated using the same SBM(0.025,0.02) graph as in Fig.~\ref{ppv_inter} and throughout. Blue, black and yellow curves respectively correspond to a realization of one of the three variant graphs described in App.~\ref{connectivities_app}. Training sets comprise $M_{train} = 2000$ paired chains AB, and in (b), testing sets include $M_{test} = 500$ paired chains AB, and were randomly partitioned into species with 10 pairs AB each. Averages over 100 realizations corresponding to different training (and different testing sets in (b)) are shown.}\label{Connectivities}
\end{figure}

For each of these graphs, we investigated the performance of both inference tasks. Because the temperature of the ferromagnetic-paramagnetic transition depends on the graph connectivity, and because performance depends on the difference between sampling temperature and this transition temperature (see App.~\ref{app_length}), we present results as a function of sampling temperature in Fig.~\ref{Connectivities}. 
%We further approximately determined the transition temperature for each of these graphs, through the slowing down of the relaxation of correlation (see Fig.~\ref{autocorrelation_fig_v2}) as well as magnetization histograms (see Fig.~\ref{histog}).

Fig.~\ref{Connectivities} demonstrates that while details depend on graph structure, the overall behavior of inference performance is the same for our two variant graphs as for our baseline graph. Note that we approximately determined the transition temperature for each of these graphs (as in Figs.~\ref{autocorrelation_fig_v2} and~\ref{histog}), and observed that its value strongly impacts the temperature where performance is highest. Therefore, our main results are robust to graph structure. Nevertheless, we observe that the maximum PPV attained for partner prediction tends to decrease when inter-block couplings are sparser. Indeed, these couplings have a smaller impact on the state of spins compared to the intra-block ones if there are fewer of them.

\clearpage

\begin{acknowledgments}
AFB thanks Ned S. Wingreen and Yaakov Kleeorin for inspiring discussions. CAGP acknowledges  Alejandro Lage, Roberto Mulet and Juan José Gonzalez Armesto for helpful discussions. PM, MW and AFB thank Institut de Biologie Paris-Seine (IBPS) at Sorbonne Université for funding via a Collaborative Grant (Action Incitative) to MW and AFB. CAGP and MW acknowledge funding by the EU H2020 research and innovation program MSCA-RISE-2016 under grant agreement No. 734439 InferNet.
\end{acknowledgments}

\end{document}